\documentclass[english,superscriptaddress,showkeys,nofootinbib]{revtex4-2}
\usepackage[T1]{fontenc}
\usepackage[utf8]{inputenc}
\usepackage{amsmath}
\usepackage{graphicx}
\usepackage[normalem]{ulem}

\usepackage[disable]{todonotes}

\usepackage{subfigure}

\usepackage{babel}

\usepackage[labeled,resetlabels]{multibib}
\newcites{S}{Bibliography}
\begin{document}
\title{Concept of Inverted Refractive-Index-Contrast Grating Mirror and Exemplary Fabrication by 3D Microprinting}

\author{Emilia Pruszyńska-Karbownik}
\affiliation{Institute of Physics, Lodz University of Technology, ul. Wólczańska 217/221, 93-005 Łódź, Poland}
\author{Daniel Jandura}
\affiliation{Department of Physics, Faculty of Electrical Engineering and Information Technology, University of Zilina, Univerzitna 1, SK-01026 Zilina, Slovakia}
\author{Maciej Dems}
\affiliation{Institute of Physics, Lodz University of Technology, ul. Wólczańska 217/221, 93-005 Łódź, Poland}
\author{Łukasz Zinkiewicz}
\affiliation{Institute of Experimental Physics, Faculty of Physics, University of Warsaw, ul. Pasteura 5, 02-093 Warsaw, Poland}
\author{Artur Broda}
\affiliation{Łukasiewicz Research Network, Institute of Microelectronics and Photonics,\\al. Lotników 32/46, 02-668 Warsaw, Poland}
\author{Marcin Gębski}
\affiliation{Institute of Physics, Lodz University of Technology, ul. Wólczańska 217/221, 93-005 Łódź, Poland}
\author{Jan Muszalski}
\affiliation{Łukasiewicz Research Network, Institute of Microelectronics and Photonics,\\al. Lotników 32/46, 02-668 Warsaw, Poland}
\author{Dusan Pudis}
\affiliation{Department of Physics, Faculty of Electrical Engineering and Information Technology, University of Zilina, Univerzitna 1, SK-01026 Zilina, Slovakia}
\affiliation{University Science Park of the University of Zilina, Univerzitna 1, SK-01026 Zilina, Slovakia;
}
\author{Jan Suffczyński}
\email{j.suffczynski@uw.edu.pl}
\affiliation{Institute of Experimental Physics, Faculty of Physics, University of Warsaw, ul. Pasteura 5, 02-093 Warsaw, Poland}
\author{Tomasz Czyszanowski}
\email{tomasz.czyszanowski@p.lodz.pl}
\affiliation{Institute of Physics, Lodz University of Technology, ul. Wólczańska 217/221, 93-005 Łódź, Poland}





\keywords{subwavelength gratings, polymer photonics, 3D microprinting}

\begin{abstract}
Highly reflective mirrors are indispensable components in a variety of state-of-the-art photonic devices. Typically used, bulky, multi-layered distributed Bragg (DBR) reflectors are limited to lattice-matched semiconductors or nonconductive dielectrics. Here, we introduce an inverted refractive-index-contrast grating (ICG), as compact, single layer alternative to DBR. In the ICG, a subwavelength one-dimensional grating made of a low refractive index material is implemented on a high refractive index cladding. Our numerical simulations show that the ICG provides nearly total optical power reflectance for the light incident from the side of the cladding whenever the refractive index of the grating exceeds 1.75, irrespective of the refractive index of the cladding. Additionally, the ICG enables polarization discrimination and phase tuning of the reflected and transmitted light, the property not achievable with the DBR. We  experimentally demonstrate a proof-of-concept ICG fabricated according to the proposed design, using the technique of 3D microprinting in which thin stripes of IP-Dip photoresist are deposited on a Si cladding. This one-step method avoids laborious and often destructive etching-based procedures for grating structuration, making it possible to implement the grating on any arbitrary cladding material.
\end{abstract}

\maketitle

\section{Introduction\label{sec:intro}}

\begin{figure*}[t]
\center\includegraphics[scale=0.64]{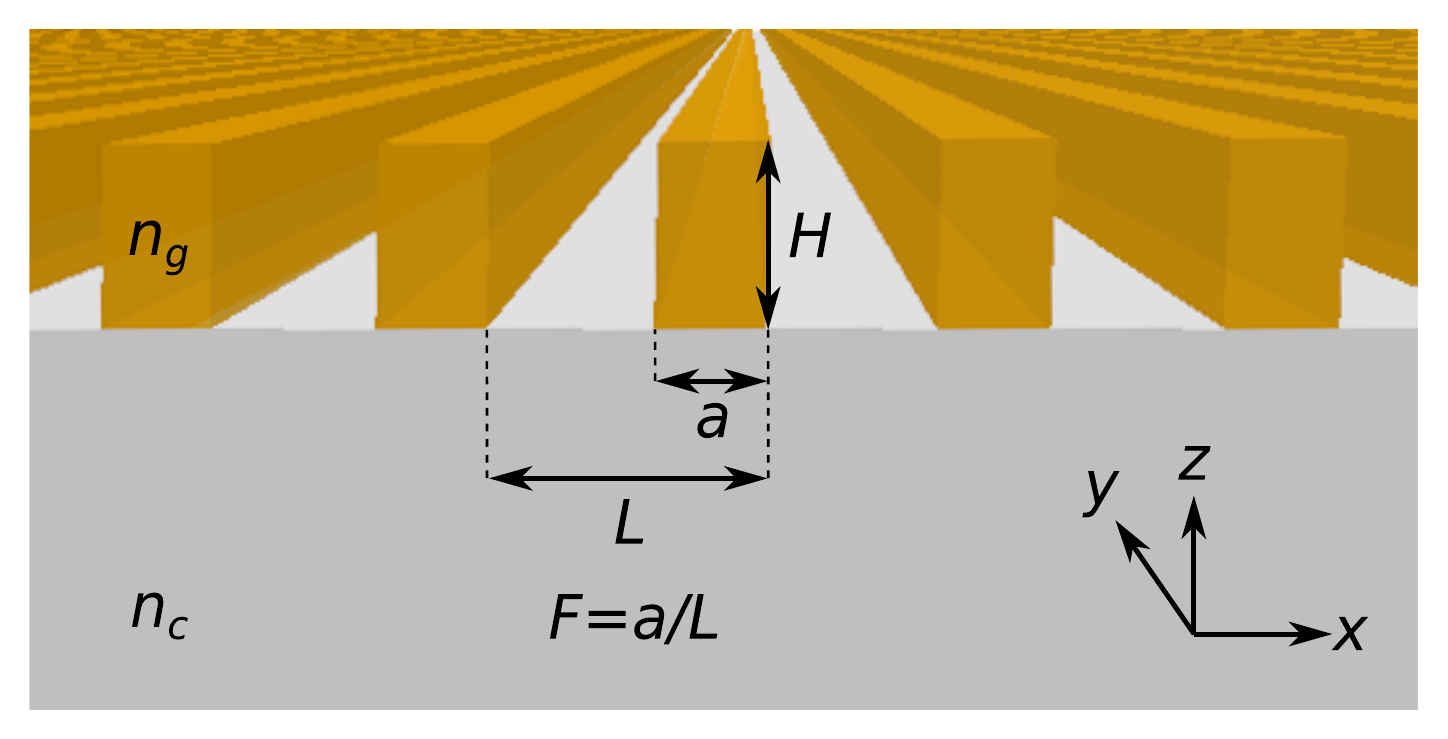}
\caption{\label{fig:structure}Configuration of the grating structure, composed of stripes with refractive index $n_g$ implemented on a cladding with refractive index $n_c$ and covered by air from the top. In the case of $n_g < n_c$ the grating is the inverted refractive-index-contrast grating (ICG). When $n_g > n_c$ the grating represents a conventional high refractive-index-contrast grating (HCG) and when $n_g = n_c$ the grating is a monolithic high refractive-index-contrast grating (MHCG). The geometrical parameters of the grating and the coordinate system are indicated.}
\end{figure*}

Reflective elements are key components in photonic and optoelectronic devices, enhancing light-matter coupling effects \cite{Kishino1991JQE,Campillo1991PRL,Bravo-Abad2007OE,Schneider2012APL,Deng2002S,Klaers2010N,Kasprzak2006N,Hess2012NM,Fas2021JPCL}. Conventional high optical power reflectance mirrors employ distributed Bragg reflectors (DBRs) composed of numerous pairs of layers with quarter-wavelength optical thicknesses and contrasting refractive indices \cite{Michalzik2013}. The epitaxial growth of most semiconductor-based DBRs is technically challenging, due to the low refractive index contrast between materials with similar lattice constants. In turn, distributed Bragg reflectors made of dielectric materials \cite{Lott1994PTL} suffer from high thermal resistivity, absence of electrical conductivity, and narrow bands of high transmission making optical pumping difficult. Optical subwavelength structures offer an attractive alternative to multilayer, several-micrometer thick DBRs. A notable example is the high refractive index contrast grating (HCG) \cite{Mateus2004PTL,Boutami2007APL}. An HCG consists of parallel, thin high-refractive-index stripes that are embedded in a low refractive index surrounding. The stripes can be placed on the top of a thick layer (Fig.~\ref{fig:structure}) made of a low refractive index material, which we call cladding \cite{Mateus2004PTL,Chung2008PTL}, or suspended in air (i.e., a striped membrane) \cite{Zhang2014LSA}. Its high optical reflection results from the destructive interference of the grating modes, which are confined due to the low refractive index of the surrounding \cite{Karagodsky2010OE}. The advantages of HCGs include their extremely high power reflectance of up to 100\% ($R = 1.0$), a broadband reflection spectrum up to two times wider spectrally than that of semiconductor DBRs \cite{Amann2008NP}, and properties that are impossible to achieve using DBRs, such as strong polarization discrimination and phase tuning of reflected light. On the other hand, the fabrication of HCGs involves a multistep procedure, as it typically relies on electron-beam lithography, photolithography, or nanoimprinting. Using these methods, a stripe-like pattern is defined in a photoresist deposited on the surface of the cladding, followed by metal deposition and lift-off to form a protective mask, before the final step of``wet'' or ``dry'' etching \cite{Oh2021JMST}.

In this paper, we introduce a new design for a highly reflecting subwavelength grating: an inverted refractive index contrast grating (ICG). The design consists of low refractive index grating stripes ($n_g$) deposited on a high refractive index cladding layer ($n_g$, Fig.~\ref{fig:structure}). Thus, the low and high refractive indices are $inverted$ ($n_g < n_c$) with respect to a conventional HCG ($n_g > n_c$).

We start with theoretical analysis indicating the possibility of high reflectance, even though the low refractive index of the grating precludes waveguiding, crucial in the case of HCGs. Subsequently, we verify the theoretical findings by numerical analysis (see Section "Numerical Methods"). For this purpose, we choose a refractive index of the cladding equal to 3.5 and consider two values for the grating refractive index: 1.5 and 2. The value of 1.5 is very low in comparison to the refractive index of materials typically implemented in reflecting metastructures. Anyway, we demonstrate a significant level of reflection and point out that the low refractive index grating is suitable for the use as a mirror in resonant cavities designed for sensing, enhancement of the spontaneous emission rate, or producing nonlinear effects. The choice of a grating refractive index of 2 is motivated by the demonstration of reflection into the zeroth diffraction order reaching nearly 100\%, which enables the realization of mirrors for a very broad range of applications in photonics and optoelectronics. We verify our theoretical and numerical analysis by comparison with experimental reflection spectra of an ICG with a very low refractive index grating, which was 3D microprinted using a photoresist polymer (IP-Dip) on silicon cladding. Microprinting is a versatile alternative to subtractive, multistep, etching-based techniques for producing 3D nanostructure and microstructures \cite{Cumpston1999N}. It has been used effectively for the fabrication of various light-harnessing structures, such as 3D photonic crystals, micro-waveguides, and micro-optical elements, including micro-lenses and miniaturized multi-lens objectives \cite{Deubel2004NM,Schumann2014LSA,Bogucki2020LSA,Gissibl2016NP, Jorg:LPR2022}.
However, there are almost no previous reports of using 3D microprinting on semiconductors to produce subwavelegth optical elements. This is due mainly to the common belief that the low refractive indices of polymers used for printing (from 1.5 to 1.58 \cite{Gissibl2017OME}), as well as the limited spatial resolution of 3D polymer microprinting, preclude the application of microprinting for the fabrication of reflecting subwavelength gratings. In fact, our work shows that 3D microprinting is very well suited for the deposition of subwavelength-scale periodic reflecting structures.

\section{Regimes of Reflection in Inverted- and High Refractive Index Contrast Gratings\label{sec:impact-of-nc}}

\begin{figure*}[ht]
\center\includegraphics[width=0.8\textwidth]{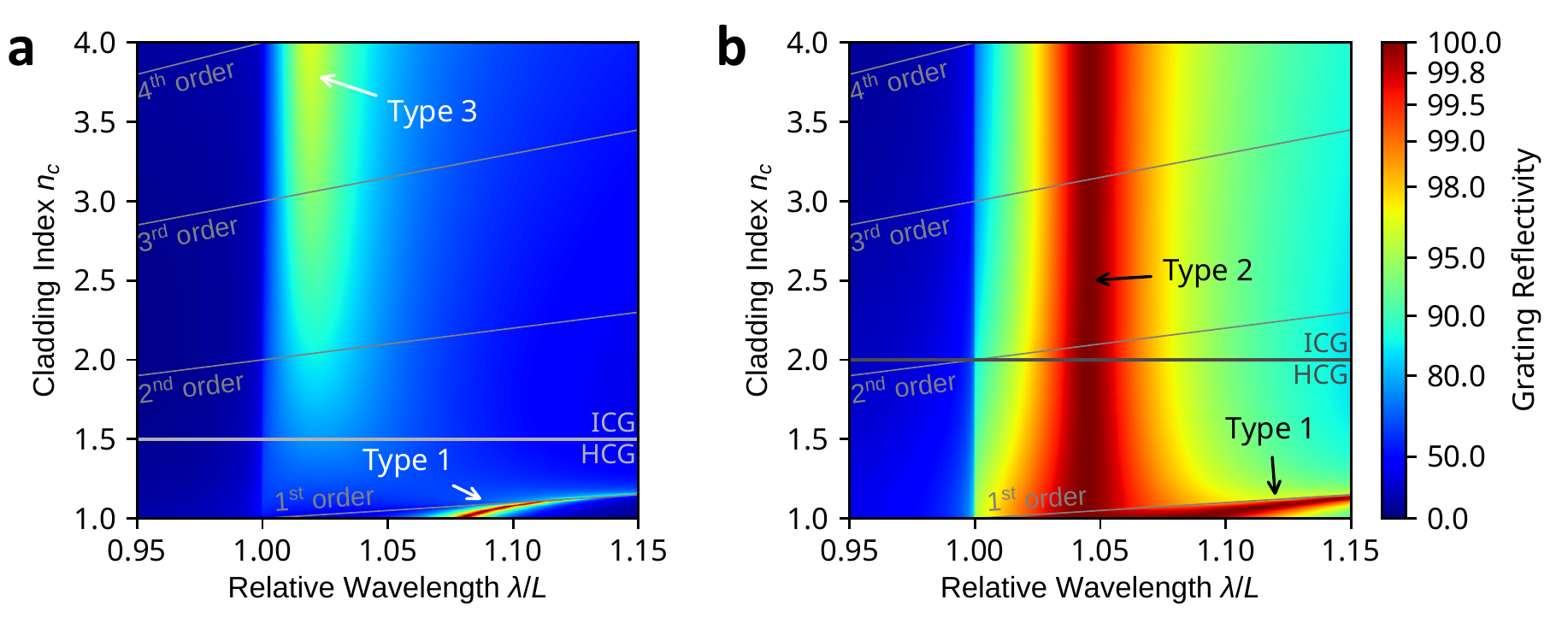}

\caption{\label{fig:icg-reflectivity-maps}Reflectivity from ICGs with refractive index $n_g = 1.5$~ a) and $2.0$~ b) calculated as a function of the relative wavelength ($\lambda/L$) and the cladding refractive index $n_c$. The other grating parameters are as follows: $H/L = 0.468$, $F = 0.395$ and $H/L = 0.373$, $F = 0.414$ for a) and b), respectively. $H$ denotes a height, $L$ -  period and $F$ - fill factor of the grating. The light incidents from the cladding (bottom) side. There are three qualitatively different mechanisms responsible for high-reflectivity peaks marked in the plots as Type-1, 2, and 3 (see text). Grey lines indicate the cut-offs of the higher-diffraction orders in the cladding.}
\end{figure*}

\begin{figure*}[ht]
\center\includegraphics[scale=0.75]{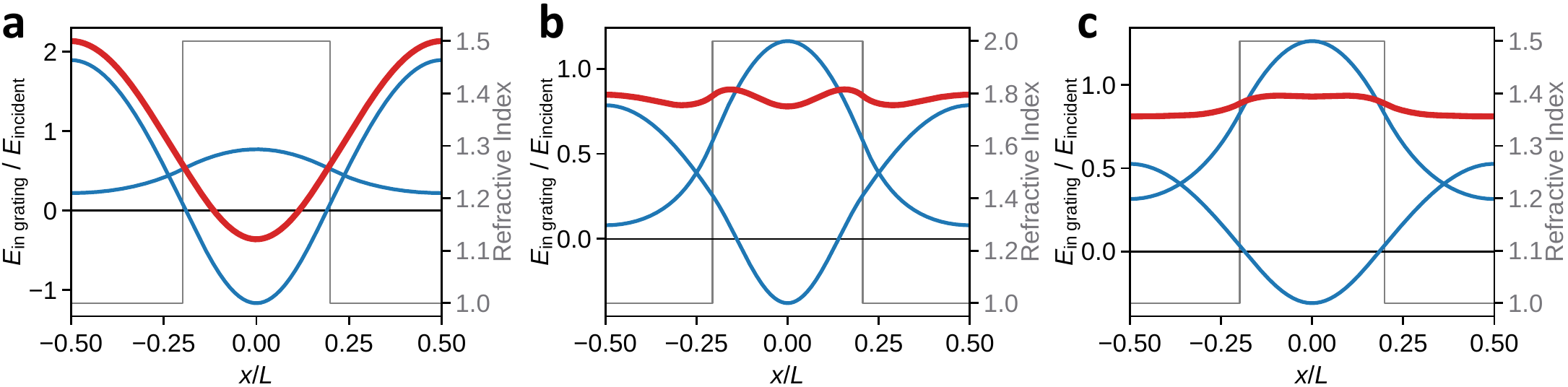}

\caption{\label{fig:grating-modes-interference}Interference of the grating modes on the input side for high reflectivity peaks Type 1, 2, and 3 in a), b), and c), respectively. The blue curves illustrate calculated profiles of the individual modes of the grating, while the red ones show their superposition. Electric field intensity is normalized relative to the one of the incident wave. For the reflection maxima of Type 2 and 3, this superposition is nearly flat, indicating a near-perfect elimination of higher diffraction orders, which is not the case for the reflection of Type 1. Gray lines indicate refractive index profile in the grating.}
\end{figure*}

In general, the high reflectivity of HCGs is a result of the two-mode interference phenomenon \cite{Karagodsky:12,Chang-Hasnain2012AOP}. In the subwavelength regime, such gratings support two modes propagating vertically (in the direction perpendicular to the grating plane) that can couple to each other only at the top and the bottom surface of the grating. Total 100\% reflection occurs when there is destructive interference of these modes on the output (top) side of the grating. On the input (bottom) side, their superposition can be arbitrary. In most conventional HCGs, high reflectivity can be obtained only when the refractive index of the cladding ($n_c$) is low enough that only a single diffraction order exists in the reflection \cite{Karagodsky:12,Chang-Hasnain2012AOP}. We name the maxima in the reflection spectrum, which are induced by this mechanism Type 1. Examples of these maxima are shown in plots in Fig.~\ref{fig:icg-reflectivity-maps} showing reflectivity maps of two different ICGs, calculated as a function of the relative
wavelength ($\lambda/L$) and the cladding refractive index $n_c$ (see Section "Numerical Methods" for details of the method). Their spectral positions depend on $n_c$ and they disappear with the appearance of higher diffraction orders in the cladding. This happens because the two grating modes interfere on the input side, such that the resulting wave couples to higher diffraction orders of the reflected wave, while the low refractive index cladding prohibits their propagation. Figure~\ref{fig:grating-modes-interference}a illustrates interference on the grating input side of a two-mode solution based on the analytical model proposed in \cite{Karagodsky:12}, for the Type-1 reflection maximum marked in Fig.~\ref{fig:icg-reflectivity-maps}a at $n_c = 1$ and $\lambda/L = 1.079$. The sine-like shape of the superposed field profile indicates strong coupling to the first and higher orders of the reflected wave.

High reflection mechanisms of a different nature are also possible, such as the maximum shown in Fig.~\ref{fig:icg-reflectivity-maps}b for the wavelength $\lambda/L = 1.046$. Based on the same analytical formalism, a single reflection channel related to the zeroth diffraction order can be identified below the first-order diffraction cut-off (Fig.~\ref{fig:icg-reflectivity-maps}b) as the only reflection channel existing in this configuration. The presence of such 100\% reflectivity spectral region is independent of the cladding refractive index. The mechanism standing behind this phenomenon is explained in the Supplementary Materials (Section~\ref{sec:S_maths}). With increasing $n_c$, higher diffraction orders of the reflected wave become possible in the cladding, but the fact that the original zero-order reflection remains unaffected by the change of $n_c$ together with the conservation of energy implies that no light is scattered into the higher-order cladding modes. An important property of this reflection mechanism, which we call Type 2, is the nearly flat superposition of the grating modes in comparison to reflection Type 1, as Figs.~\ref{fig:grating-modes-interference}a and~\ref{fig:grating-modes-interference}b illustrate. In this mechanism, the zero-order component of the grating mode Fourier expansion dominates significantly over other components, enabling 100\% reflectivity into the zeroth diffraction order.

The last type of high reflectivity mechanism, named Type 3, is responsible for a reflection peak in Fig.~\ref{fig:icg-reflectivity-maps}a for the wavelength $\lambda/L = 1.022$. As will be demonstrated in Section \ref{sec:numerical}, Type 3 may appear when the refractive index of the grating ($n_g$) is less than 1.75, whereas for such low $n_g$ Type 2 is absent. Type 3 provides less than 100\% reflectivity into the zeroth diffraction order, as the interference of the modes is not fully destructive. However, the reflected light propagates almost solely in the zeroth diffraction orders, due to the accidental flattening of the superposition of the grating modes at the input side, as shown in Fig.~\ref{fig:grating-modes-interference}c. 

The theoretical analysis presented in this section demonstrates the physical mechanism responsible for the emergence of very high reflectivity in the case of the subwavelength grating of refractive index lower than that of the cladding. In the following section we characterize the impact of ICG geometry on optical properties using numerical approach.

\section{Numerical Verification of Inverted Refractive Index Contrast Gratings Properties\label{sec:numerical}}

In this section we numerically calculate the power reflectance of a grating with the refractive index $n_g$ deposited on the surface of the semi-infinitely thick monolithic cladding with the refractive index $n_c$ larger than $n_g$ ($n_g < n_c$). A semi-infinite air superstrate is assumed above the grating (see Fig.~\ref{fig:structure}). In the calculations, we consider reflection in the zeroth-diffraction order only and a single period of the grating with periodic boundary conditions, which elongates the grating to infinity in the lateral direction. The normal incidence of the light from the cladding side is assumed. 
As a reference, we also consider the case of an HCG where $n_g > n_c$, as well as the border case of a monolithic HCG (MHCG), where $n_g = n_c$ \cite{Chang-Hasnain2012AOP,Gebski2015OE}.

\begin{figure*}[tbh]
\center\includegraphics[width=1\textwidth]{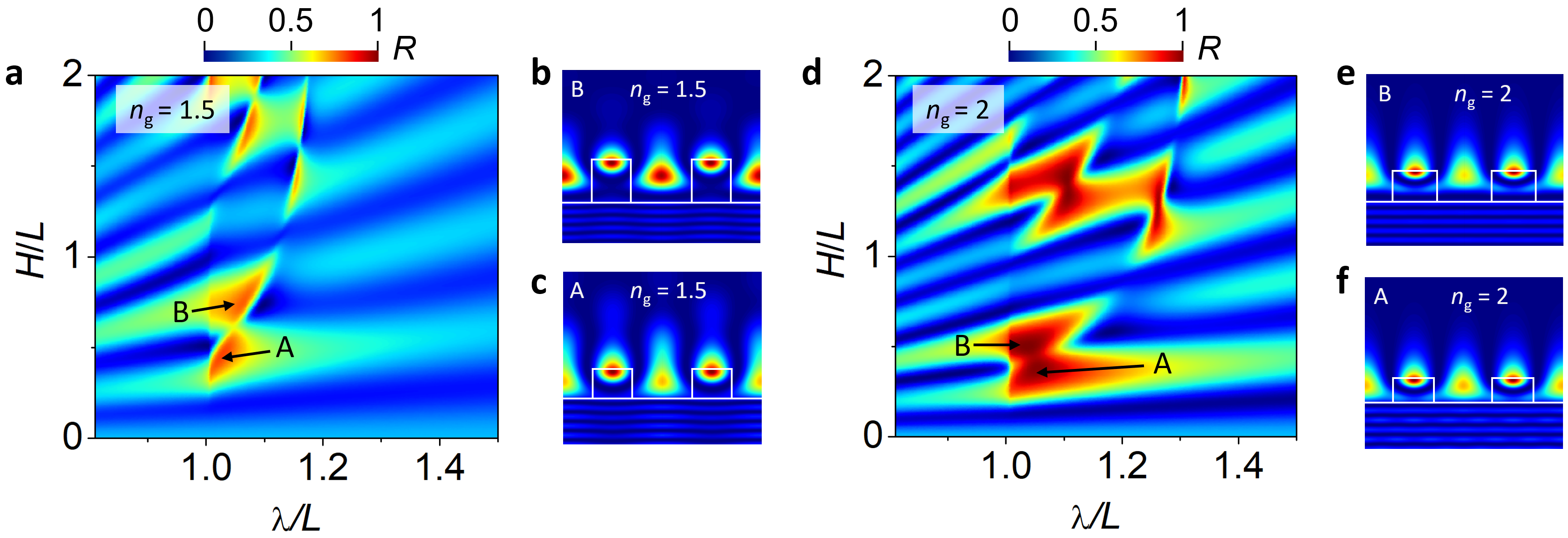}
\caption{\label{fig:reflect_map_1}Calculated reflectance map for the inverted contrast grating (ICG) in the domain of wavelength $\lambda$ and grating height $H$, both relative to the grating period $L$. The refractive index of the cladding $n_c$ is 3.5. The refractive index of the grating and fill factor are assumed as $n_g = 1.5$, $F = 0.395$ in a) and $n_g = 2.0$, $F= 0.414$ in d).
In b), c), e), f) the distributions of optical field intensity corresponding to the \textit{A} and \textit{B} reflection maxima for $n_g = 1.5$ b), c) and $n_g = 2.0$ e), f) are shown within an ICG illuminated by a plane wave at normal incidence from the cladding side. The parameters of \textit{A} and \textit{B} maxima are collected in Table \ref{tab:geometry}.}

\end{figure*}

Figures \ref{fig:reflect_map_1}a, d show maps of power reflectance for two exemplary ICGs. Calculations are conducted in the domain of the height of the stripes ($H$) and the light wavelength ($\lambda$). We consider grating material with a refractive index $n_g = 1.5$ (see Fig. \ref{fig:reflect_map_1}a) or $n_g = 2.0$ (see Fig. \ref{fig:reflect_map_1}d) and the same cladding layer ($n_c = 3.5$) in both cases.

The ICG modes are leaky (see Sections \ref{sec:impact-of-nc}), whereas HCG grating modes are not; however, the reflection pattern visible in the maps resembles to some extent the ``checkboard'' pattern observed also in the case of HCG \cite{Chang-Hasnain2012AOP}. As shown in Supplementary Materials \ref{sec:S_numerIntegrICG} this region is limited by cutoffs of $\mathrm{TE}_{2n}$ (from the short wavelength side) and the long-wavelength limit, according to waveguide theory \cite{Snyder1994}. Above the long-wavelength limit, only $\mathrm{TE}_{0n}$ modes exist and the grating behaves as a quasi-uniform (unstructured) layer. Therefore, the reflection resembles a Fabry-Perot interference pattern, produced by a uniform layer without any regions of high power reflectance.

\begin{table}[tbh]
\caption{\label{tab:geometry}Geometrical parameters of the inverted refractive-index-contrast grating (ICG) for configurations corresponding to \textit{A} and \textit{B} maxima for $n_g = 1.5$ and $n_g = 2.0$: $L$ – period of the grating, $F$ – fill factor, $H$ – height of the stripe, $R$ – optical power reflectance into the 0$^\mathrm{th}$ diffraction order for $n_c = 3.5$.}
\center\begin{tabular}{ccccc}
\hline
maximum &\textit{A} & \textit{B} & \textit{A} & \textit{B}\\
\hline
$n_g$&1.5&1.5&2.0&2.0\\
$\lambda/L$&
1.022&
1.057&
1.046&
1.032\\
$F$&
0.395&
0.397&
0.414&
0.447\\
$H/L$&
0.468&
0.755&
0.373&
0.502\\
$R$&
0.840&
0.810&
0.996&
$1-1.8\cdot 10^{-4}$\\

\hline
\end{tabular}
\end{table}

Several regions of high power reflectance are visible in both reflection maps (Fig. \ref{fig:reflect_map_1}a, d), confirming the predictions of the theoretical model presented in Sections \ref{sec:impact-of-nc}. In what follows, we focus on the two power reflectance maxima (PRM) that we name \textit{B} and \textit{A} (Figs. \ref{fig:reflect_map_1} and Table \ref{tab:geometry}). They feature the smallest height and the broadest width of reflection stopband (WRS), which we define as a reflection stopband above 60\%. Therefore they appear to be the most attractive configurations for real-world applications.

In the case of the ICG with $n_g = 1.5$, \textit{A} and \textit{B} PRMs reach more than 80\% (Fig. \ref{fig:reflect_map_1}a). Both PRMs are located near the mode $\mathrm{TE}_{20}$ (see Supplementary Fig. \ref{fig:fields}), which influences the optical field distributions in the grating, contributing to a single optical field maximum along the $z$ axis in the region of the grating as illustrated in Figs. \ref{fig:reflect_map_1}b, c. Increasing the grating refractive index to $n_g = 2.0$ increases the grating reflectance to nearly 100\% for \textit{A} and \textit{B} PRMs and broadens their WRS, as illustrated in Fig. \ref{fig:reflect_map_1}d. Light distributions corresponding to \textit{A} and \textit{B} PRMs in the case of $n_g = 2.0$ are illustrated in Figs. \ref{fig:reflect_map_1}e, f displaying similar light distribution as in the case of $n_g = 1.5$. The geometrical parameters of ICG configurations corresponding to \textit{A} and \textit{B} PRMs for $n_g = 1.5$ and $n_g = 2.0$ are collected in Table \ref{tab:geometry}. The reflection spectra corresponding to the four maxima are presented in Supplementary Fig. \ref{fig:spectral_S}.

Closer inspection of light distributions for the PRMs for $n_g = 1.5$ and $n_g = 2.0$ reveals that the dominant maximum intensity of the light is located close to the top surface of the stripe. Moreover, the optical field extends into the air above the stripe, independently of its refractive index \cite{Kravtsov2020LSA,Zhang2018NC}. There is also a significant build-up of light density in the grating (see Supplementary Materials \ref{sec:S_numerIntegrICG}) that may be possibly utilized to enhance light-matter interaction in the region of the grating, as demonstrated in \cite{Zhang2018NC}. Light distribution for the \textit{B} maximum for $n_g = 1.5$ shows additional significant local maxima in the air slit between the stripes (Fig. \ref{fig:reflect_map_1}b), which could facilitate interaction of the reflected light with the surroundings, enabling possible sensing applications \cite{Sun2016SR} in proximity to the ICG.  Further properties of ICGs are discussed in more details in the Supplementary Materials \ref{sec:S_ref_spect_phase} here we indicate the most important conclusions. First concerns possibility of high transmission of the light incident from the air side. This property together with the very high reflectance of the zeroth diffraction order when light is incident from the cladding side, is expected to be useful when the ICG constitutes one or both mirrors of a Fabry-Perot cavity subjected to external excitation. Another property of the ICG is possibility of phase tuning of reflected light at the level of $d\phi/d\lambda \approx 10\pi \mathrm{rad}$, which provides a facile method of tuning the resonant wavelength of a cavity with an ICG mirror, by modifying the geometrical parameters of the ICG while keeping the cavity thickness constant \cite{Haglund2016OE}.

\begin{figure*}[tbh]
\center\includegraphics[width=0.4\textwidth]{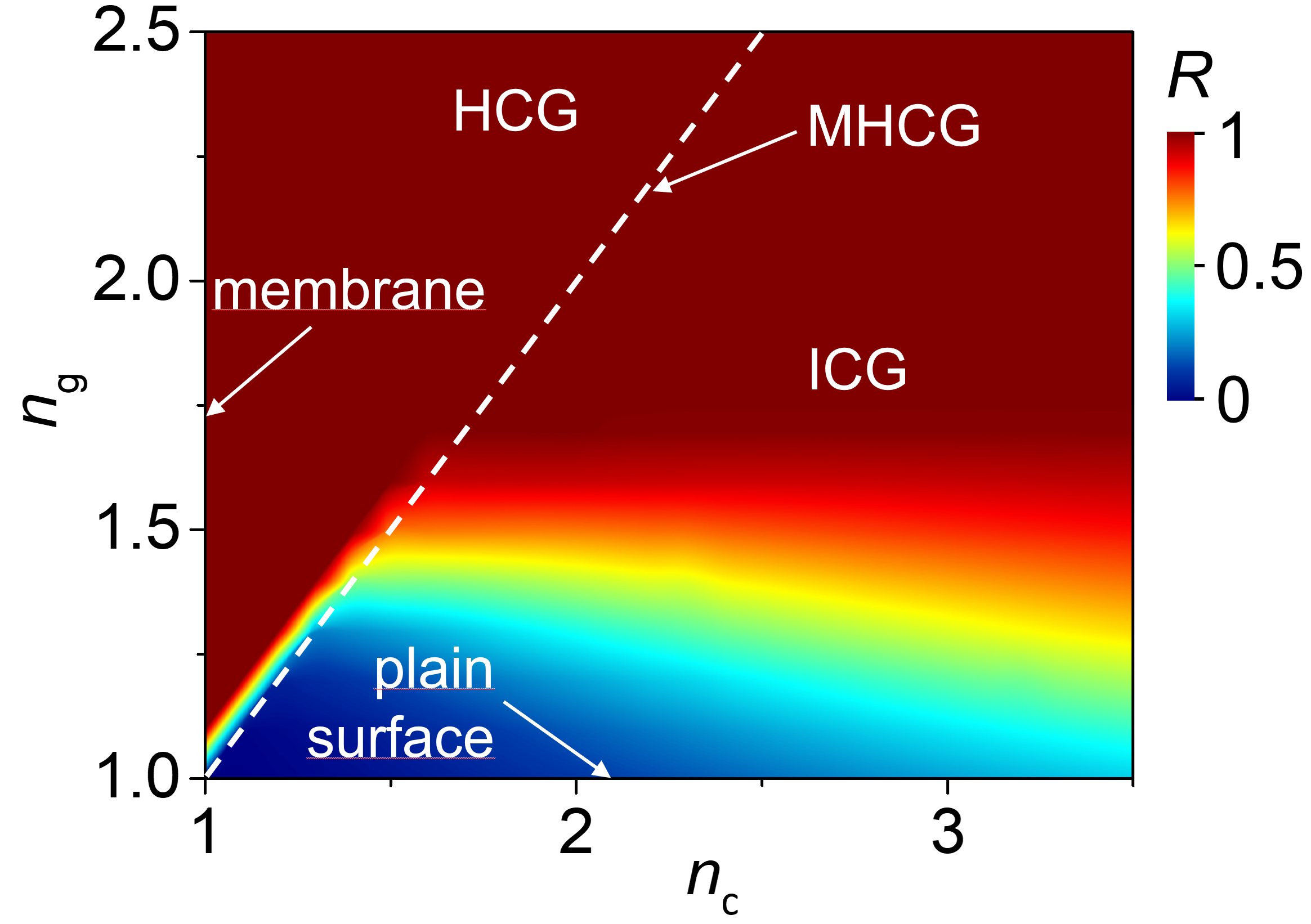}
\caption{\label{fig:max}Map of maximal power reflectance for subwavelength gratings calculated in the ($n_c$, $n_g$) space. Each point on the map represents maximal reflection of \textit{B} and \textit{A} PRMs. The white dashed line represents the MHCG configuration ($n_c = n_g$). The region positioned on the left of the dashed line represents an HCG ($n_c < n_g$). The region on the right of the dashed line represents an ICG ($n_c > n_g$). The vertical line at $n_c = 1$ represents a membrane suspended in air. The horizontal line at $n_g = 1$ corresponds to a plain surface between the cladding and the air.}
\end{figure*}

A more general picture of the optical performance of the ICG and all possible subwavelength grating configurations is provided in Fig. \ref{fig:max}, showing the calculated maximal power reflectance of the gratings in the domain of $n_g$ and $n_c$ for light incident from the cladding side. Magnitude of each point on the map is the largest value for either the \textit{B} or \textit{A} reflection maximum (see Supplementary Materials \ref{sec:S_ref_spect_ng}). The geometrical parameters of the gratings are modified throughout the map, since modifying the refractive index of grating imposes different conditions for the optimal geometrical parameters ensuring the maximal reflectance. Power reflectance of 100\% into the zeroth diffraction order is achieved by all HCG configurations that fulfil the condition $n_g > n_c$, including the membrane configuration in which the grating is suspended in air ($n_c = 1$). The MHCG configuration ($n_g = n_c$) enables total reflectance when the refractive index of the grating and the cladding is larger than 1.75, in agreement with Ref. \cite{Marciniak2016OL}. A previously unexplored feature is an apparent ability of the ICG to achieve nearly 100\% reflection when $n_g < n_c$, which is related to Type 2 reflection as discussed in Section \ref{sec:impact-of-nc}. The only requirement is that $n_g$ is larger than 1.75. For $n_g > 1.75$, the total reflection is found within numerical precision as long as the difference $n_c-n_g$ is less than 0.5, while for $n_c-n_g > 0.5$ the maximal power reflectance into the zeroth diffraction order is not smaller than $1-10^{-3}$. With a decrease in the refractive index of the grating ($n_g < 1.75$), the power reflectance and WRS also decrease revealing features of Type 3 reflection (see Section \ref{sec:impact-of-nc}). However, as shown in Fig. \ref{fig:max}, an ICG with $n_g < 1.75$ still provides power reflectance considerably exceeding the reflectance of the plain surface between the cladding and air. The influence of the refractive index of the grating $n_g$ on the reflection spectrum of ICG is analysed in more detail in the Supplementary Materials \ref{sec:S_ref_spect_ng}.

As discussed above, an HCG in which $n_g > n_c$ ensures total reflection into the zeroth diffraction order and a wide WRS. However, this configuration requires the implementation of a high refractive index material, such as a semiconductor, on a lower refractive index thick layer, for example a dielectric. This impedes the use of such mirrors in resonant optoelectronic devices, including vertical-cavity surface-emitting lasers, due to practical problems with current injection, heat dissipation, and mechanical stability compared with all-semiconductor configurations. Combining two semiconductor layers of different refractive indices to achieve an HCG is also demanding, due to the typically significant difference in the lattice constants of semiconductor materials with sufficiently high refractive index contrast. Eliminating these problems is possible, in principle, by using an MHCG; however, the fabrication of MHCGs remains a challenge. The concept of an ICG in which a lower refractive index grating is deposited on cladding with a higher refractive index can substantially simplify grating implementation, due to the considerable freedom of forming thin dielectric subwavelength structures on top of semiconductor devices. In particular, the dielectric-semiconductor boundary can be a natural etch-stop that enables better control over the etched structure parameters. An ICG composed of semiconductor cladding and a dielectric grating fabricated using dielectrics with refractive indices higher than 1.75, such as TiO$_2$ (refractive index in the range of 2.05--2.48 \cite{Sarkar2019AMI}), TaO$_2$ (2.08--2.3 \cite{Gao2012OE}), or Si$_3$N$_4$ (1.98--2.05 \cite{Luke2015OL}), would allow for nearly 100\% reflection into the zeroth diffraction order of the normal incident light from the semiconductor side. If materials of even lower refractive index, such as SiO$_2$ (1.4--1.5 \cite{Kischkat2012AO}) or IP-Dip photoresist (1.5--1.58 \cite{Gissibl2017OME}) are deposited on an arbitrary semiconductor, reflection into the zeroth diffraction order is expected to reach 85\% and nearly 98\% into all diffraction orders, as will be demonstrated in the next section.

\section{Experimental Demonstration of 3D Microprinted ICG\label{sec:experiment}}
To experimentally verify the theoretical model and numerical simulations presented in Sections \ref{sec:impact-of-nc} and \ref{sec:numerical}, we fabricated an ICG  using 3D microprinting technique. The low refractive index of IP-Dip $n_g  = 1.53$ \cite{Gissibl2017OME} prevents 100\% power reflectance into the zeroth diffraction order. However, expected reflectance above 80\% is required in a variety of optical and optoelectronic applications, including resonator cavity enhanced light emitting diodes and resonant cavity enhanced photodetectors. Uniquely, 3D microprinting enables flexibility in the fabrication of the ICG, making possible wavelength, phase, and wavefront tuning by tailoring the parameters of the ICG stripes. Figure \ref{fig:SEM} and Fig. \ref{fig:SEM_balwan_real} in the Supplementary Materials
illustrate an ICG fabricated by 3D microprinting directly on a thick Si wafer with a refractive index of $n_c = 3.5$ at a wavelength of 1500 nm.

\begin{figure*}[tbh]
\center\includegraphics[width=0.5\textwidth]{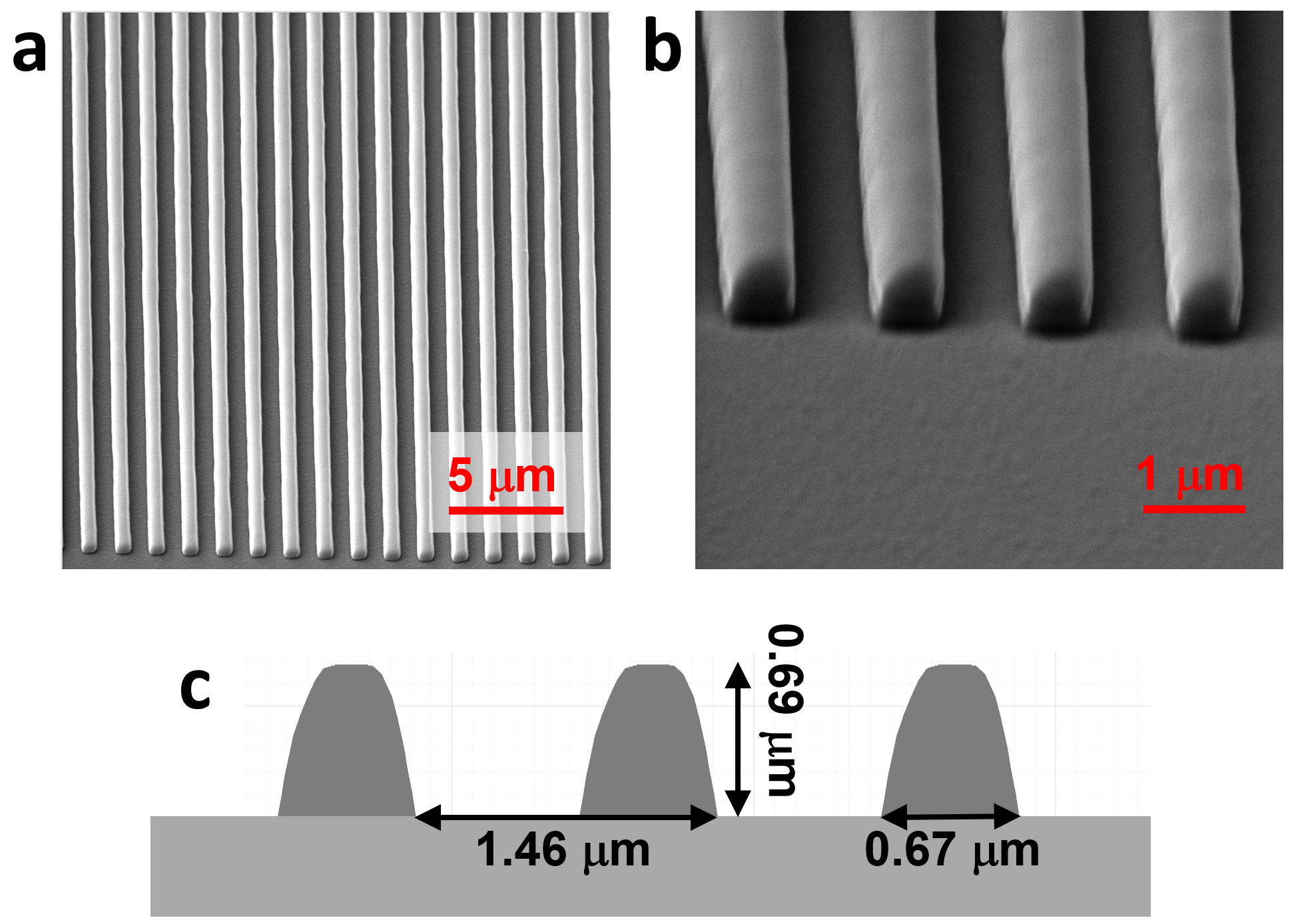}
\caption{\label{fig:SEM}a), b) Scanning electron microscope (SEM) image of a 3D microprinted IP-Dip grating deposited on Si cladding in two consecutive enlargements; c) profile of the ICG stripes scanned from SEM images and implemented in the software with the grating dimensions indicated.}
\end{figure*}

The ICG was designed for peak TE-polarized reflection at $\lambda= 1500$ nm. The double side polished Si wafer was covered with an antireflective Si$_3$N$_4$ coating, consisting of a single quarter-wavelength thick layer on the surface opposite to the surface on which the ICG was implemented. The ICG was designed with the following parameters: $\lambda/L= 1.022$, $F = 0.4$, $H/L = 0.47$, and $L = 1460\,\mathrm{nm}$, corresponding to \textit{A} PRM in Fig. \ref{fig:reflect_map_1}a. The parameters were predicted to provide maximal reflectance, assuming a rectangular cross-section of ICG stripes. The process of grating fabrication is detailed in the section Fabrication Methods.

The actual geometrical dimensions of the processed ICGs were determined by scanning electron microscopy (SEM), and $H$ was additionally inspected using a confocal optical microscope. For the presented sample, $L = 1460\,\mathrm{nm}$ (determined with 50 nm precision; see Supplementary Materials \ref{sec:S_calc_real}), $F = 0.45$ and $H/L = 0.46$.
To validate the numerical analysis, the actual cross sections of the ICG stripes were extracted from the SEM images (see Fig. \ref{fig:SEM}c). The obtained profiles were implemented in the numerical model. In what follows, all numerical results relate to the cross-section shape of the real-world ICG. Reflection maps calculated for the ICG (see Fig. \ref{fig:balwan_cyfr}a in the Supplementary Materials) show great similarity to the reflection maps of an ICG consisting of stripes with a rectangular cross section (see Fig. \ref{fig:reflect_map_1}a). The deviation in the cross-section in our experiment from the rectangular shape does not affect maximal reflection, but in general it may require modification of the grating parameters to achieve maximal power reflectance \cite{Marciniak2021ACSP}. The power reflectance of the ICG with the real-world cross section is discussed in more detail in the Supplementary Materials \ref{sec:S_calc_real}.

The transmission through the ICG sample can be expressed as follows:
\begin{subequations}
\label{eq:transmissions}
\begin{align}
T_{\mathrm{ICG}}= T_{\mathrm{ar}}  e^{-\alpha d} \left( 1-R_{\mathrm{ICG}} \right)\\
T_{\mathrm{ref}}= T_{\mathrm{ar}}  e^{-\alpha d} \left( 1-R_{\mathrm{plain}} \right)
\end{align}
\end{subequations}
where $T_{\mathrm{ICG}}$ is transmission measured for normal incident light through the ICG, $T_{\mathrm{ref}}$ is the reference transmission through the neighboring unprocessed plain silicon surface, $T_{\mathrm{ar}}$ is transmission through the antireflecting coating (which also accounts  for any scattering occurring in the wafer and on its surface), $\alpha$ is the absorption coefficient of the silicon wafer, $d$ is the thickness of the wafer, and $R_{\mathrm{plain}}$ is the reflection from the plain interface between Si and air (which is 0.304 at the wavelength of 1500 nm based on Fresnel equations for reflection).
With the set of equations (\ref{eq:transmissions}) the reflectivity of the ICG ($R_{\mathrm{ICG}}$) can be extracted directly from the transmission measurements:
\begin{equation}
\label{e:R_ICG}
R_{\mathrm{ICG}}=1-\frac{T_{\mathrm{ICG}}}{T_{\mathrm{ref}}}\left(1-R_{\mathrm{plain}}\right)
\end{equation}

\begin{figure*}[tbh]
\center\includegraphics[width=0.8\textwidth]{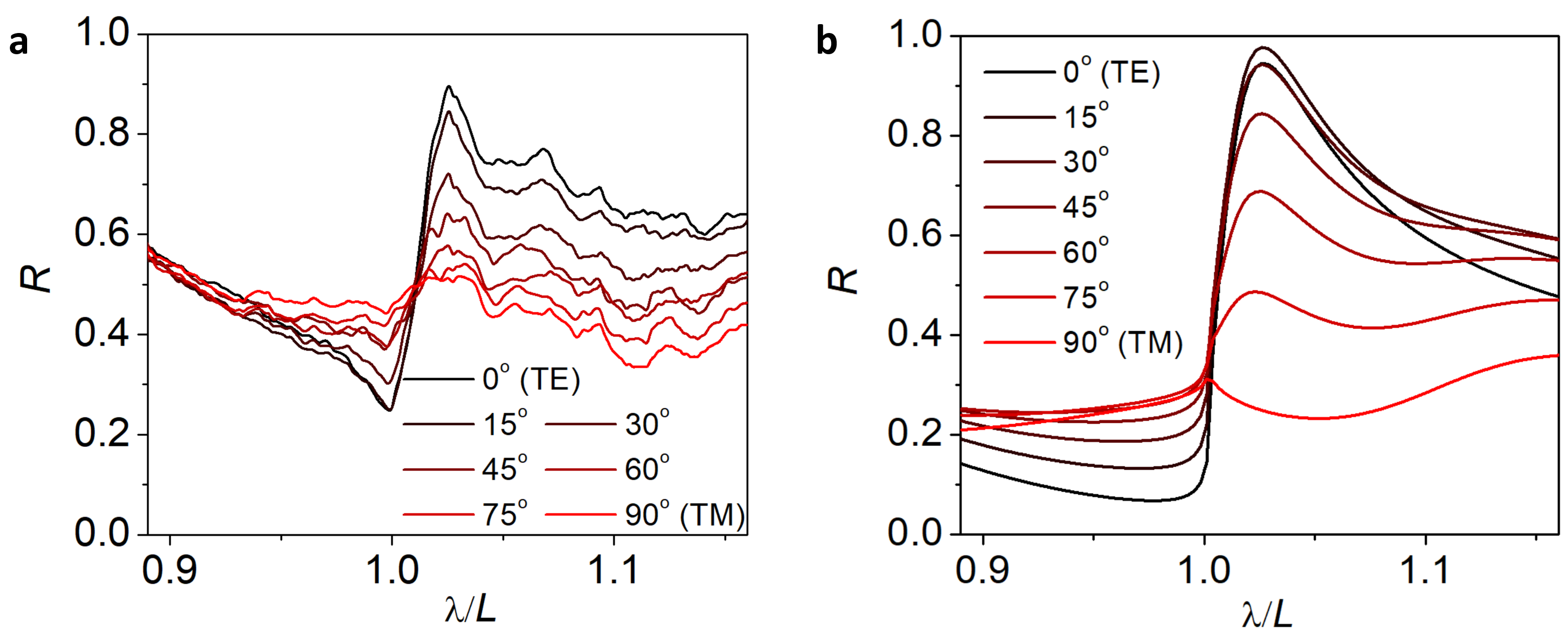}
\caption{\label{fig:experiment}Measured a) and calculated b) reflection spectra into all diffraction orders of inverted refractive-index-contrast gratings (ICGs) 3D microprinted using IP-Dip on silicon cladding. The spectra are calculated for the geometrical parameters of the stripes determined experimentally, with the grating defined by the following parameters: $L = 1460\,\mathrm{nm}$, $F = 0.45$, $H/L = 0.46$. The spectra are measured and calculated for polarization modified gradually with a 15-degree step from TE to TM. The spectra of TE and TM polarization correspond to angles of the polarizer of 0 and 90 degrees, respectively. }
\end{figure*}

Figure \ref{fig:experiment} presents the measured (Fig. \ref{fig:experiment}a) and calculated (Fig. \ref{fig:experiment}b) reflection spectra for various angles of the polarizer rotating from 0 to 90 degrees with a 15 degree step. The extreme angles of rotation represent TE and TM polarisations.
In the measurements, only the zeroth diffraction order was transmitted through the grating, due to the dimensions of the subwavelength stripes. Therefore, $R_\mathrm{ICG}$ accounts for all diffraction orders of light reflected by the grating. The experimental reflection spectra for TE polarisation reveal a local maximum at $\lambda/L= 1.027$ ($\lambda \approx 1500\,\mathrm{nm}$) that corresponds very well with the numerical results. The measured maximal reflection is close to 90\%, which is also close to the numerical simulations, revealing maximal reflection into all diffraction orders at the level of 97\%. The TE reflection abruptly reduces towards shorter wavelengths, which is consistent with the calculated reflection map in Fig. \ref{fig:reflect_map_1}a, indicating high transmission in this spectral range. Rotation of the polarizer from the position corresponding to TE polarization to the position corresponding to TM polarization reduces the reflectivity to nearly 30\% that is a level of a reflectance from the interface between silicon and air . At the wavelength corresponding to the maximal reflectance, TE polarisation reflection is twofold larger than TM polarisation reflection. The calculations show that TE polarization reflection can be fivefold larger than TM polarization reflection. The inconsistencies in the measurements and simulations are typically related to the fabrication precision, which introduces deviations in the grating periodicity. The experimental power reflectance can be enhanced by perfecting the process of ICG fabrication. Overall, our experimental results show very good agreement with calculations and confirm the feasibility of high power reflectance using an ICG.

\section{Conclusions\label{sec:conclusions}}

We have presented a new high reflecting mirror design for an inverted refractive index contrast grating, along with a theoretical, numerical, and experimental investigation of its optical performance. By theoretical analysis we demonstrated the possibility of high reflectance independent of the refractive index of the cladding on which the grating is deposited, particularly when refractive index of the grating is lower than the refractive index of the cladding.

By numerical analysis, we showed that the ICG provides almost 100\% optical power reflectance for light incident at normal from the cladding layer side toward the air. The only requirement is for the grating to be formed from a material with a refractive index higher than 1.75. The refractive index of the layer below the grating can be arbitrary. When the refractive index of the grating is less than 1.75, the grating still strongly enhances the power reflectance compared to reflection occurring at the plane interface between the cladding layer and air. The ICG enables polarization control of reflected light, with fivefold larger reflection of transverse electric (TE) polarization compared to transverse magnetic (TM) polarization, and facilitates phase tuning of reflected light.

To experimentally verify our numerical analysis, we characterized the optical reflectance of an IP-Dip grating fabricated by 3D microprinting on a thick silicon wafer. Qualitative and quantitative comparison of the measured and calculated power reflectance spectra revealed very good agreement, indicating nearly 90\% reflection into all diffraction orders and strong polarization control.

At a more general level, the proposed design and its implementation using an additive-type technique open up a new possibilities for the fabrication of subwavelength structures, which are in increasing demand in photonics, optics, and optoelectronics. The fabrication of highly reflective mirrors in the form of 3D microprinted gratings does not require high-vacuum techniques such as vapor deposition or epitaxy, and has the additional advantage of scalability. Thanks to the relaxation of the requirements for the refractive index of the cladding layer hosting the grating, the range of materials that can be applied is extended, making the use of perovskite or organic grating layers possible.

\section*{Numerical methods\label{sec:num_methods}}

To determine the optical reflectance of the gratings, we use the plane-wave reflection transformation method \cite{Dems2011OER}, which is a fully vectorial optical model. Because of the periodicity of the gratings, the electrical field of the electromagnetic wave can be expressed in the form of Bloch waves: $\Psi(x)=e^{ik_x x}f(x)$, where $f(x)$ is a periodic function with the same period as the grating $L$, and $k_x$ is the lateral component of the wavevector of the light, ranging from $-\pi/L$ to $\pi/L$. In the analysis, we use 60 plane waves that enable numerical relative error below $10^{-8}$. The model has been shown to have high reliability by comparison with experimental results \cite{Jandura2020ASS,Marciniak2020OE}. In the analysis we consider transverse electric (TE) polarization, where the electric field is parallel to the grating stripes. Transverse magnetic (TM) polarization perpendicular to the grating is not considered here, as the ICG shows significantly lower power reflectance of this polarization.

\section*{Fabrication methods\label{sec:fab_methods}}

To fabricate the ICG grating, we used the Photonic Professional GT laser lithography system from Nanoscribe GmbH with a $63 \times$ immersion objective and IP-Dip polymer material. The system uses Er-doped femtosecond frequency-doubled fiber laser emitting pulses at 780 nm wavelength with an approximately 100~MHz repetition rate and 150~fs pulse width. The femtosecond laser is focused into the volume of the IP-Dip photoresist, where the two-photon polymerization process occurs in the volume of the focal spot (voxel). In the fabrication process, the IP-Dip polymer was deposited on top of the silicon substrate and polymerization by laser writing was realized layer by layer in a single-step process. The grating structure on top of the silicon cladding was fabricated using a programmed script in a two-layer arrangement of horizontal stripes, with laser power of 26~mW and a scanning speed of 10000\,\textmu m/s. For the development of a polymerized structure, PGMEA (propylene glycol monomethyl ether acetate) was applied for 20~min to dissolve and remove the unexposed photoresist. Finally, the sample was rinsed in isopropyl alcohol for 4~min and dried with nitrogen.

\section*{Measurements}
For the transmission measurements, a supercontinuum light source (Leukos SM-30-W; 400--2400 nm) was coupled to the optical fiber, illuminating the sample from the side. The polarizer for 1550 nm was placed between the supercontinuum source and the sample and a rotary stage was used to change the angle of polarization. On the opposite side of the sample, a detection optical fiber was moved precisely toward the sample using an immersion layer. The transmission spectra of the ICG grating were measured using an OceanOptics NIRQuest spectrometer (900--2050~nm) with respect to the reference transmission of the silicon substrate.

\begin{figure*}[tbh]
\center\includegraphics[width=0.5\textwidth]{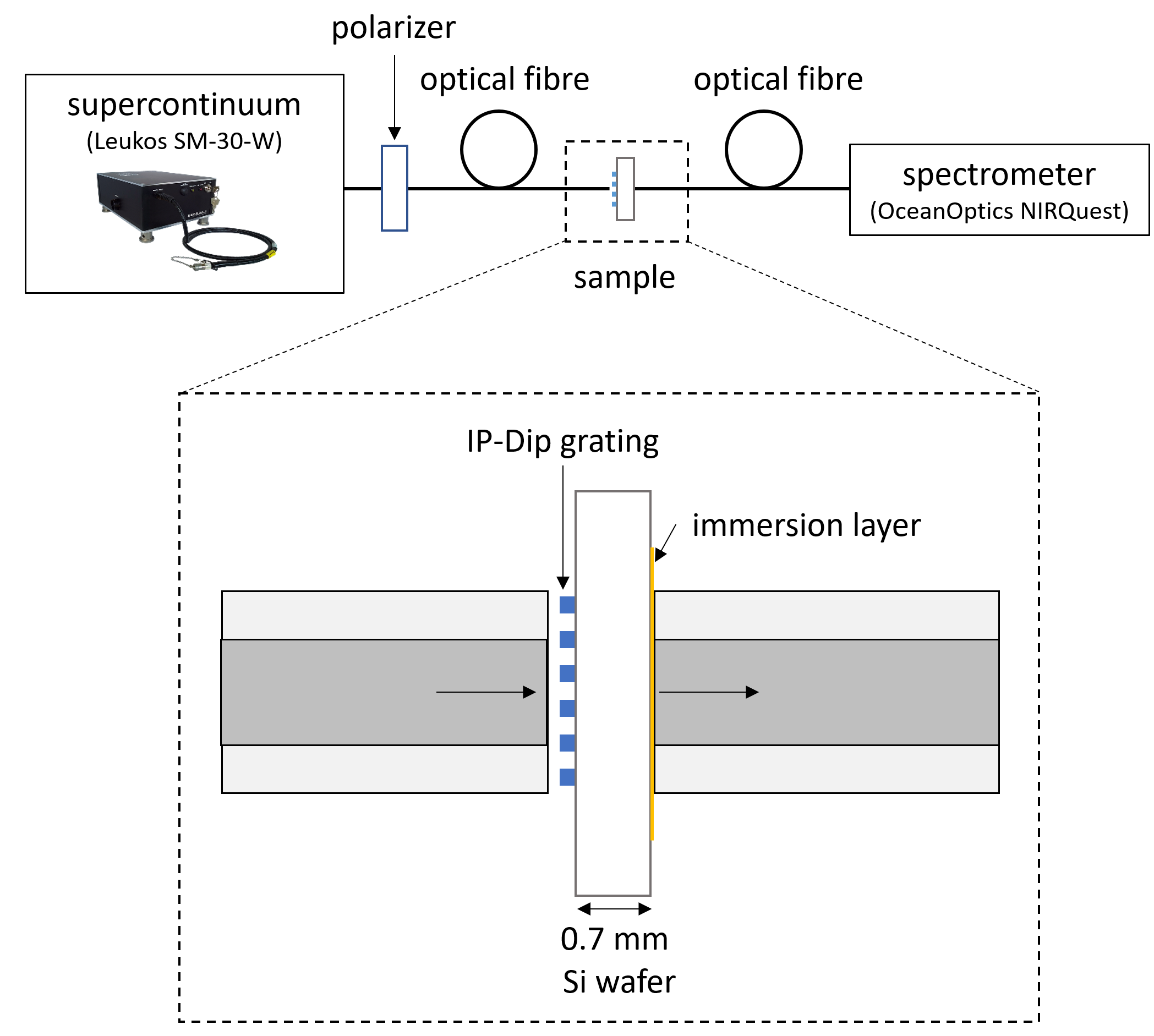}
\caption{\label{fig:set-up}Diagram of the experimental setup.}
\end{figure*}

\section*{Acknowledgements}
This work is supported by the Polish National Science Center within the projects OPUS 2018/29/B/ST7/01927, 2017/25/B/ST7/00437 and 2020/39/B/ST7/03502 and by the Slovak National Grant Agency under project No. VEGA 1/0363/22.

\bibliographystyle{naturemag}
\bibliography{ihcg}

\newpage
\setcounter{page}{1}
\setcounter{figure}{0}
\setcounter{table}{0}
\setcounter{section}{0}

\renewcommand{\thepage}{S\arabic{page}}
\renewcommand{\thesection}{S\arabic{section}}
\renewcommand{\thetable}{S\arabic{table}}
\renewcommand{\thefigure}{S\arabic{figure}}

\renewcommand{\citenumfont}[1]{S#1}

\part*{Supplementary Materials}
\renewcommand{\theequation}{S\arabic{equation}}
\setcounter{equation}{0}

This PDF file includes supporting:

Information \ref{sec:S_maths} -- \ref{sec:S_calc_real}

Figures \ref{fig:compar} -- \ref{fig:balwan_cyfr}

Table \ref{tab:geometry}

\newpage

\section{\label{sec:S_maths}Explanation of the properties of Type 2 and Type 3 reflectivity}

The peculiar behavior of the Type-2 reflection peak---its independence from the refractive index of the cladding---is the consequence of the relation between the electric and magnetic field at the reflection side of the grating\,\citeS{S_Dems2017JLT}. Let us consider a zero-order TE-polarized plane wave incident from the substrate side, traveling into a perfect infinite grating. It is sufficient to consider only single components of the electric and magnetic fields parallel to the grating ($E_{y}$ and $H_{x}$). The normalized fields in the cladding at the grating boundary are
\begin{subequations}
\begin{align}
E^{(1)}(x,z) & =1+\sum_{n}r_{n}\cos(2\pi nx/L)\mbox{,}\\
H^{(1)}(x,z) & =\alpha_{0}-\sum_{n}r_{n}\alpha_{n}\cos(2\pi nx/L)\mbox{,}
\end{align}
\end{subequations}
where $n$ is the number of the diffraction order, $r_{n}$ is the amplitude reflection coefficient of the $n$-th diffraction order, and $\alpha_{n}$
is
\begin{equation}
\alpha_{n}=\sqrt{\varepsilon_c-\left(n\lambda/L\right)^{2}}\mbox{,}\label{eq:alpha}
\end{equation}
where $\varepsilon_c=n_c^{2}$ is the cladding permittivity.

To remove $x$-dependence from the above equations, they are multiplied by $\cos(2\pi mx/L)$ and integrated over the grating period. This allows vectors of consecutive orders of the electric ($\mathbf{e}$) and magnetic ($\mathbf{h}$) fields to be represented as
\begin{subequations}
\begin{align}
\mathbf{e} & =\mathbf{d}_{0}+\mathbf{r}\text{,}\\
\mathbf{h} & =\boldsymbol{\alpha}\left(\mathbf{d}_{0}-\mathbf{r}\right)\text{,}
\end{align}
\end{subequations}
where $\mathbf{d}_{0}=\left[\begin{array}{cccc} 1 & 0 & 0 & \ldots\end{array}\right]^{T}$ is Kronecker's delta vector, $\mathbf{r}$ is the vector of the reflection coefficients, and $\boldsymbol{\alpha}$ is the diagonal matrix with elements defined by Eq.~(\ref{eq:alpha}). It is possible to derive a relation between $\mathbf{e}$ and $\mathbf{h}$ using only grating parameters, independently of the cladding refractive index in the form\,\citeS{S_Dems2017JLT}
\begin{equation}
\mathbf{h}=\mathbf{Z}\thinspace\mathbf{e}\text{.}
\end{equation}
The matrix $\mathbf{Z}$ is the input impedance matrix. This yields the reflection of the system as
\begin{equation}
\mathbf{r}=\left(\boldsymbol{\alpha}+\mathbf{Z}\right)^{-1}\left(\boldsymbol{\alpha}-\mathbf{Z}\right)\mathbf{d}_{0}\text{.}\label{eq:r}
\end{equation}

For the Type-2 high reflectivity peak, the diagonal elements of the impedance matrix dominate over non-diagonal elements. Furthermore, the zero-order element $Z_{00}$ is purely imaginary. In such case, the zero-order reflection coefficient is (from Eqs.~(\ref{eq:alpha}) and~(\ref{eq:r}))
\begin{equation}
r_{0}\approx\frac{\varepsilon_c-Z_{00}}{\varepsilon_c+Z_{00}}\mbox{.}
\end{equation}
As $Z_{00}$ is purely imaginary, $|r_{0}|\cong1$ regardless of the value of $\varepsilon_c$. In other words, the zero-order reflectivity coefficient is always 100\% regardless of the value of $n_c$, which only impacts the phase of the reflected wave. In accordance with the law of conservation of energy, all other reflected diffraction orders are suppressed.

Analogous analysis for reflection maxima related to Type 3 indicate non-negligible $|r_{0}|$ and $|r_{1}|$ diagonal elements. This indicates the existence of zeroth and first diffraction orders in the reflection, precluding 100\% reflection into the zeroth diffraction order.

\newpage

\section{\label{sec:S_numerIntegr}Numerical determination of modes dispersion by the light confinement in the grating}

\begin{figure*}[tbh]
\center\includegraphics[width=0.9\textwidth]{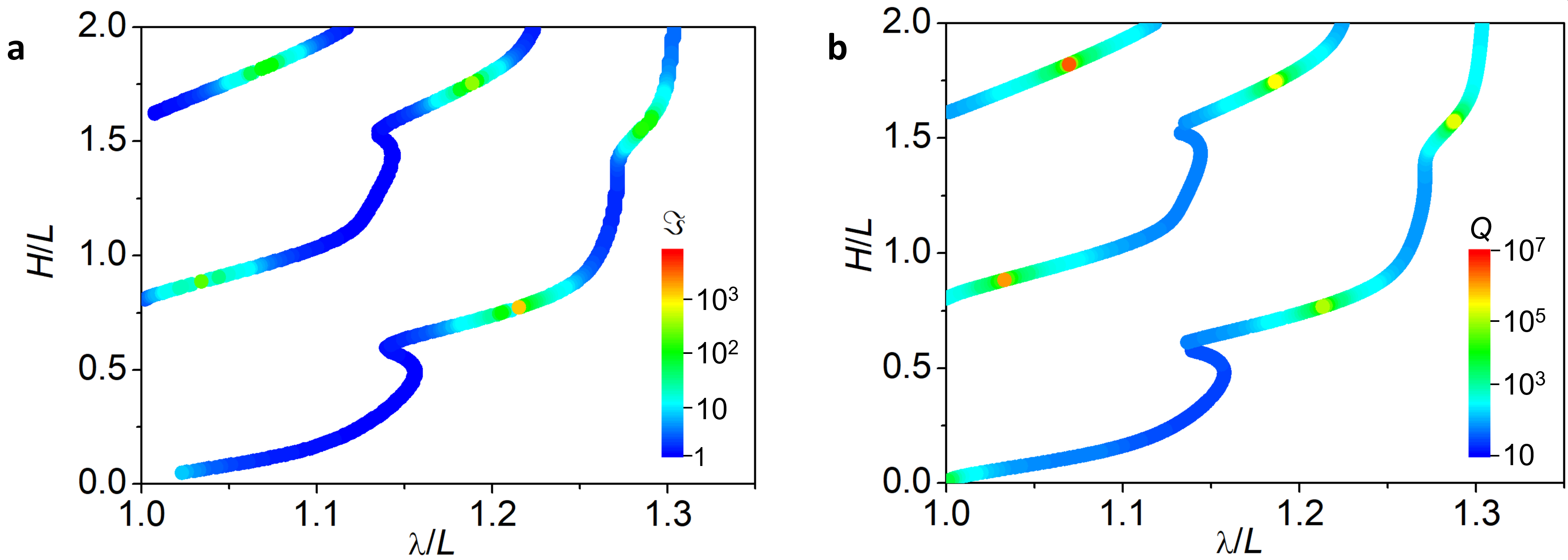}
\caption{\label{fig:compar}Mode dispersion in the domain of wavelength $\lambda$ and grating height $H$ both relative to the grating period $L$ calculated by PWRTM in a) and PWAM in b). The refractive index of the grating $n_g$ is 2 and refractive index of the surroundings is 1. The fill factor $F$ of the grating stripes is 0.4. Colors represent relative light confinement $\Im$ in ICG in a) and $Q$-factor in b).}
\end{figure*}

In this section and the section that follows we consider dependence of the resonant wavelength of the modes on the height of the grating stripes and this dependence for brevity is called dispersion. Numerical identification of the dispersion curves of the grating modes using an eigenvalue solver based on the plane wave admittance method (PWAM) \citeS{S_Dems:05} is significantly hindered by the very low $Q$-factor of the leaky modes in an ICG, which become untraceable for the solver. Therefore, we identify the modes using the method described in \citeS{S_Marciniak2021ACSP}, whereby the modes are identified by the build-up of light intensity inside the ICG as minima according to the formula
\begin{equation}
D=\frac{\partial^{2}}{\partial H^{2}}\Im=\frac{\partial^{2}}{\partial H^{2}}\frac{H_{c}\int_{\mathrm{ICG}}EE^{*}\,dy\,dz}{H\int_{\mathrm{c}}EE^{*}\,dy\,dz}
\end{equation}
where $E$ is the electric component of the electromagnetic field that is determined by the plane-wave reflection transformation method (PWRTM) described in the "Numerical methods" section in the main text. The integral in the nominator is over the ICG layer with thickness $H$ and the integral in the denominator is over the cladding layer with $H_c$, which is significantly larger than the wavelength in the cladding. By $\Im$ is designated the relative light confinement in the ICG. Figure \ref{fig:compar} compares the dispersion curves of $\mathrm{TE}_{20}$, $\mathrm{TE}_{21}$, and $\mathrm{TE}_{22}$ modes in a grating membrane composed of stripes with a refractive index of 2 surrounded by air. In such a membrane configuration, the quality ($Q$) factor of the modes is high enough to be traceable by PWAM and therefore enables direct comparison with the method based on $\Im$ determination by PWRTM. The dispersion curves determined by PWRTM in Fig. \ref{fig:compar}a and by PWAM in Fig. \ref{fig:compar}b have no perceptible differences. The points composing the curves display colours corresponding to $\Im$ in Fig. \ref{fig:compar}a and to $Q$-factor in Fig. \ref{fig:compar}b. Local maxima of the $Q$-factor relate to interference-based bound states in the continuum (BIC) occurring in vertically symmetric gratings \citeS{S_Hsu2016}. Symmetry-protected BICs are absent in the figures, due to the boundary conditions in PWAM enabling laterally symmetric modes only that emit zeroth diffraction order under normal direction. Therefore normal incidence considered in the PWRTM enables coupling the light to the symmetric grating modes only. The positions of the local maxima of $Q$ and $\Im$ are also located for the same grating parameters, which results from the fact that an increase in the $Q$-factor accompanies an increase in the density of the optical field inside the cavity.

\newpage

\section{\label{sec:S_numerIntegrICG}Numerical determination of leaky modes dispersion by the light confinement in the ICG}

\begin{figure*}[tbh]
\center\includegraphics[width=0.9\textwidth]{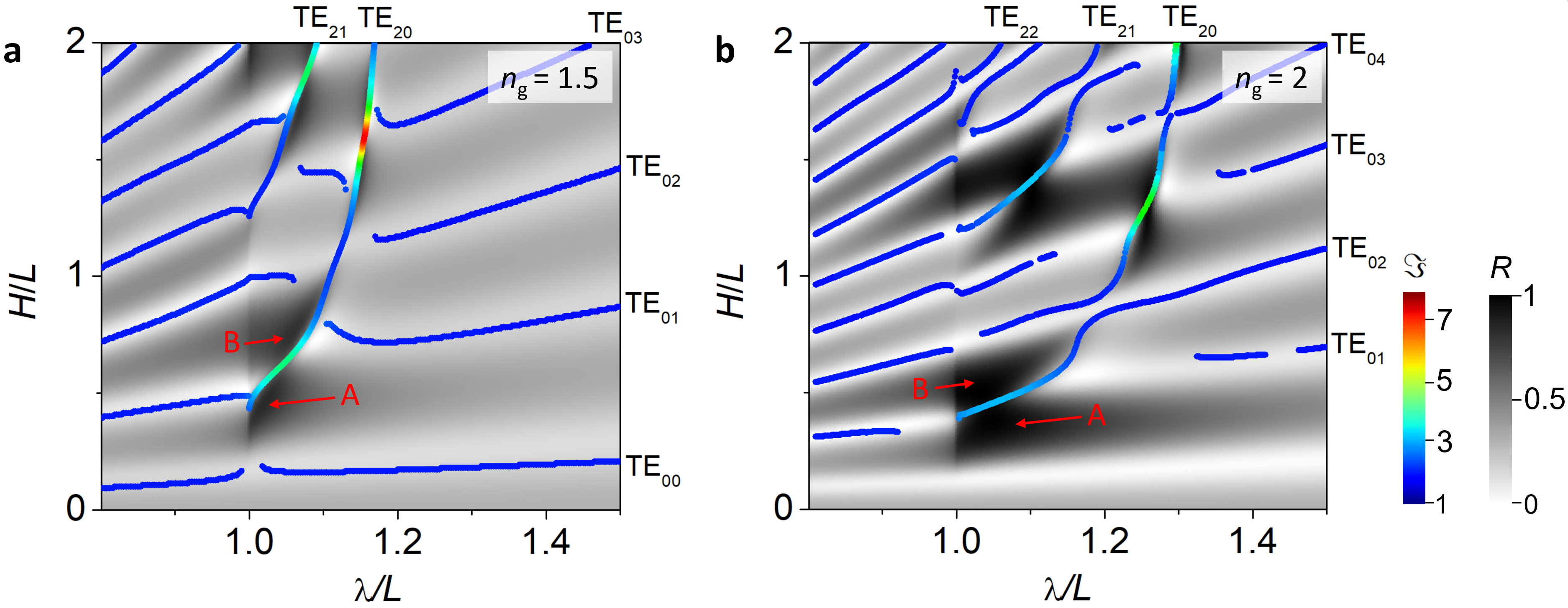}
\caption{\label{fig:fields}Reflectance ($R$) maps displayed in grey scale for the inverted contrast grating (ICG) in the domain of wavelength $\lambda$ and grating height $H$ both relative to the grating period $L$. The refractive index of the cladding $n_c$ is 3.5. The refractive index of the grating and fill factor are $n_g = 1.5$, $F = 0.397$ in a) and $n_g = 2.0$, $F= 0.447$ in b). The colour lines in a) and b) represent the dispersion of ICG modes and their relative light confinement in the ICG $(\Im)$ as defined in Section \ref{sec:S_numerIntegr}. Positions of reflection maxima \textit{A} and \textit{B} are indicated by arrows.}
\end{figure*}

Figures \ref{fig:fields} demonstrate the reflectivity maps of ICG configurations that are also presented in Fig. \ref{fig:reflect_map_1} in main text with overlapped dispersion of leaky modes existing in the structures calculated by the method described in section \ref{sec:S_numerIntegr}. PRMs \textit{A} and \textit{B} are positioned in the proximity of the dispersion of $\mathrm{TE}_{20}$ mode that affects the light distribution in the case of both maxima.
We also determine the relative light confinement $\Im$ of the modes that is indicated by color. In the case of $\mathrm{TE}_{20}$ mode there is a significant build-up of light density in the grating reaching an 8-fold increase for $n_g = 1.5$ and a 5-fold increase for $n_g = 2.0$ compared to the light density in the cladding.

\newpage

\section{Phase of reflected light and polarization discrimination\label{sec:S_ref_spect_phase}}

\begin{figure*}[tbh]
\center\includegraphics[width=0.7\textwidth]{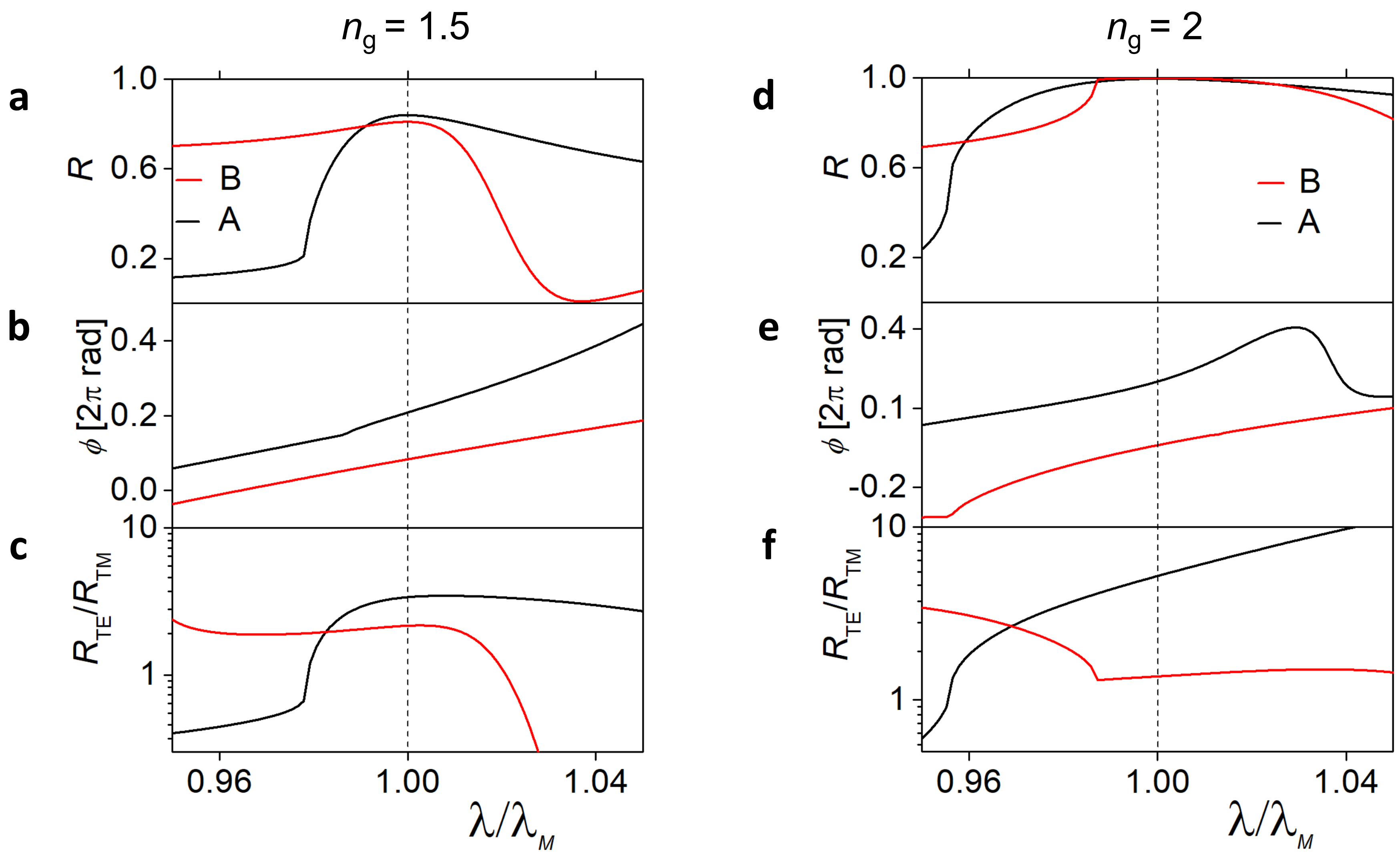}
\caption{\label{fig:spectral_S}Spectral dependences of a), d) reflection, b), e) phase of the reflected light, c), f) ratio of optical power reflectance between TE and TM polarizations, for inverted refractive index contrast grating (ICG) configurations corresponding to \textit{B} and \textit{A} PRMs indicated by red and black, respectively. Details of the configurations are summarized in Table \ref{tab:geometry}. Wavelength is unitless and $\lambda=1$ represents the central wavelength corresponding to maximal optical power reflectance.}
\end{figure*}

Figure \ref{fig:spectral_S} summarizes the principal properties characterizing the performance of the considered gratings for normal incidence of light. The spectrum shown in Fig. \ref{fig:spectral_S}d for an ICG with $n_g = 2$ indicates maximal power reflectance of nearly 100\%. The reflectance decreases to 85\% when $n_g$ decreases to 1.5, as illustrated in Fig. \ref{fig:spectral_S}a. Figures \ref{fig:spectral_S}b and e show that the considered gratings ensure efficient tuning of the phase ($\phi$) of the reflected light. In both cases, corresponding to $n_g = 1.5$ and $n_g = 2$, $d\phi /d\lambda \approx 10\pi$ rad at the vicinity of the reflection maximum. This property provides a facile method of tuning the resonant wavelength of a cavity with an ICG mirror, by modifying the geometrical parameters of the ICG while keeping the cavity thickness constant \citeS{S_Haglund2016OE}. Figures \ref{fig:spectral_S}c and f demonstrate the ratio of reflectance of TE to TM polarized light. The level of polarisation discrimination is similar to that achieved previously with HCGs and MHCGs \citeS{S_Chang-Hasnain2012AOP,S_Gebski2019OE,S_Marciniak2020OE,S_Hong2021PR}. Several times larger power reflectance for TE polarization with respect to TM could allow strong polarization discrimination of the stimulated emission when an ICG is applied as a mirror in a Fabry-Perot cavity.

\newpage

\section{Dispersion of reflection and transmission\label{sec:S_ref_disp}}

\begin{figure*}[tbh]
\center\includegraphics[width=0.8\textwidth]{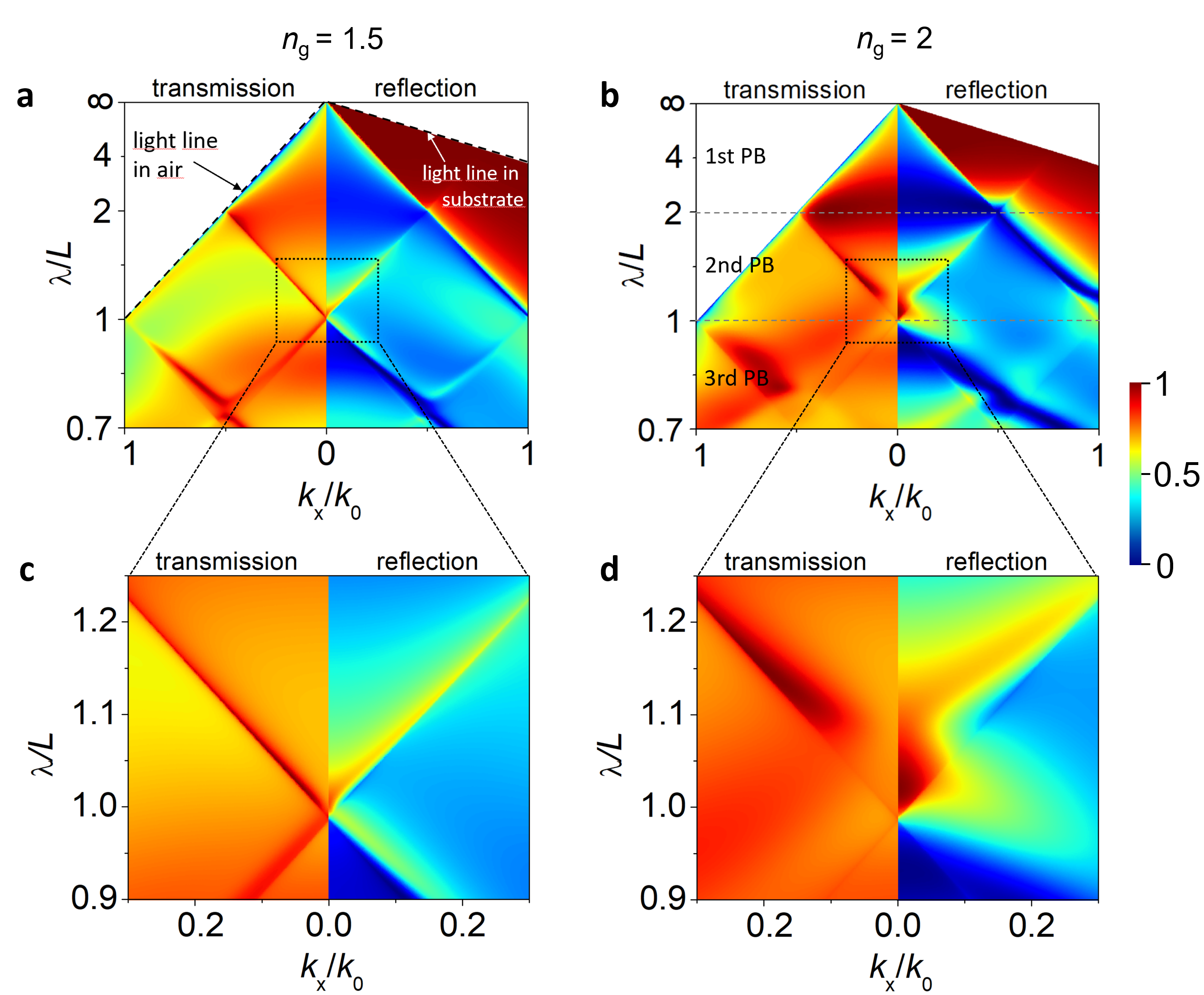}
\caption{\label{fig:spectral_1}Dispersion diagrams of transmission (left panels) and reflection (right panels) for configurations of inverted refractive index contrast gratings defined by a grating refractive index $n_g = 1.5$, fill factor $F = 0.395$, stripe height $H/L = 0.468$ in a), c) corresponding to configuration (1) in Fig. \ref{fig:reflect_map_1}a in the main text and a grating refractive index of $n_g = 2.0$, fill factor $F = 0.414$, stripe height $H/L = 0.373$ in b), d) corresponding to configuration (3) in Fig. \ref{fig:reflect_map_1}b in the main text.
In a) the borders between the photonic bands (PB) are indicated with horizontal dashed lines. In b) the light lines in the cladding and air are indicated with dashed lines. The rectangular black dotted lines in a) and b) indicate the borders of the regions illustrated in c) and d)}.
\end{figure*}

To characterize the dispersion of transmission and reflection, we calculate both values as functions of $\lambda/L$ and $k_x/k_0$, where $k_x$ is the $x$ component (see Fig. \ref{fig:structure} in the main text) of the incident photon momentum and $k_0$ is the wavenumber of the light in a vacuum. Transmission and reflection are shown in the left and right panels of the subplots in Fig. \ref{fig:spectral_1}. The maps illustrate a selected range of the plane ($k_x/k_0$, $\lambda/L$), in the vicinity of $k_x = 0$ corresponding to normal incidence. The junction of the second and third photonic bands (PB) lies in the center of the first Brillouin zone \citeS{S_Joannopoulos2008}.
The reflection is calculated for the zeroth diffraction order for the incidence from the cladding side, whereas the transmission is calculated for all diffraction orders and for light incident from the air side. This choice is motivated by the fact that in the case of a vertical cavity with an ICG as a mirror, only the zeroth diffraction order of the reflected light is involved in Fabry-Perot resonance, whereas in the case of external optical excitation the total transmission of light through the mirror may contribute to excitation of the active material embedded in the cavity. As can be seen in Fig. \ref{fig:spectral_1}, high reflectivity closely follows the folded branch of the light line in the subwavelength regime.
Comparison of the reflection panels in Figs. \ref{fig:spectral_1}a and \ref{fig:spectral_1}b indicate that the high reflectivity region widens with respect to $k_x$ when $n_g$ increases from 1.5 to 2. The presence of an abrupt transition between high and low reflectivity regions for $n_g = 1.5$ suggests that the high reflection is due to Fano-like resonance. A build-up of light intensity is also observed in Fig. \ref{fig:reflect_map_1}c in the main text, which supports the hypothesis that Fano resonance is responsible for the strong reflection in the case of gratings with a lower refractive index. Analysis of the transmission panels indicates that the transmission of the ICG is relatively high and uniform in the analyzed range of $k_0$ and $k_x$ values. It is additionally enhanced in proximity to the folded branches of the light line.

\newpage

\section{Reflection into all diffraction orders and power reflectance spectra for arbitrary $n_g$\label{sec:S_ref_spect_ng}}

\begin{figure*}[b!]
\center\includegraphics[width=1\textwidth]{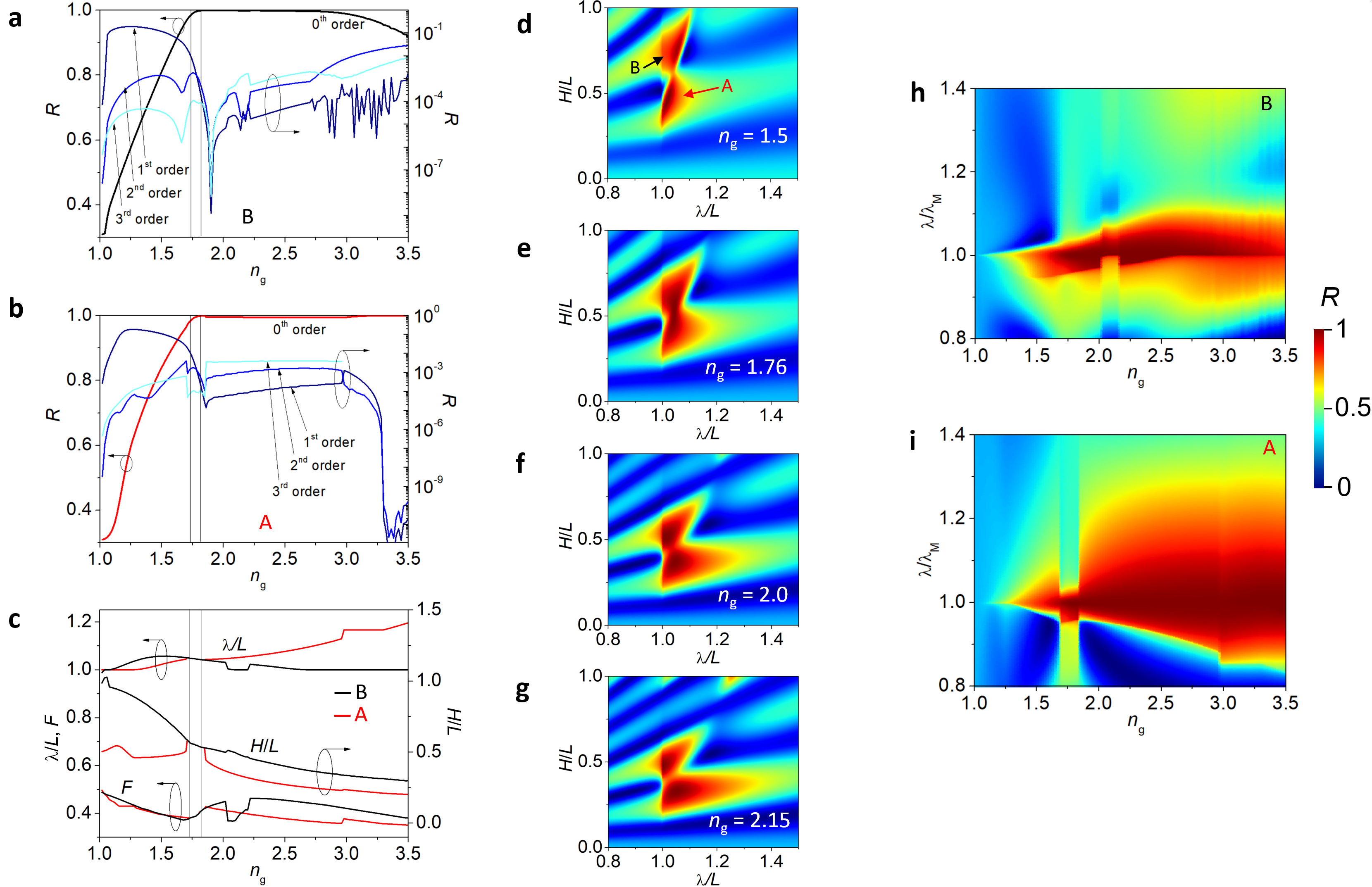}
\caption{\label{fig:refl_map_S}
ICG power reflectance into all diffraction orders as functions of the refractive index of the ICG grating $n_g$ in the case of \textit{B} a) and \textit{A} b) reflection maxima indicated in d); c) wavelength $\lambda$, height $H$ (both normalized by period $L$) and fill factor $F$ of the ICG stripes as functions of the refractive index of the grating $n_g$ for \textit{B} (black) and \textit{A} (red) reflection maxima. Values of $\lambda/L$ and $F$ are indicated on the left vertical axis while $H/L$ is indicated on the right axis. In d)--g) are shown reflectance maps in the domain of $\lambda/L$ and $H/L$ for different refractive indices of the ICG grating $n_g$,
map of power reflectance spectra for $n_g$ varying from 1 to 3.5 for \textit{B} h) and \textit{A} i) reflection maxima, where $\lambda_{M}$ is the wavelength corresponding to the reflection maximum. The refractive index of the cladding $n_c$ is 3.5 in all presented cases.}
\end{figure*}

Figure \ref{fig:refl_map_S} illustrates ICG reflection for variable $n_g$ and $n_c = 3.5$ in the cases of two local reflection maxima for the smallest $H$. The local maxima are found using multidimensional optimization, including $L$, $F$, and $H$ of the ICG as variables. Slowly varying the positions of \textit{B} and \textit{A} maxima with respect to the grating parameters when $n_g$ is modified ensures that both maxima can be tracked carefully. The geometrical parameters of the maxima for \textit{A} and \textit{B} are indicated with red and black lines, respectively, in Fig. \ref{fig:refl_map_S}c. Maximum \textit{B} (Fig. \ref{fig:refl_map_S}a) exhibits above 0.99 power reflectance into the zeroth diffraction order in the range of $n_g$ from 1.75 to 3.0 and total reflection above $1 - 10^{-4}$ in the range of $n_g$ from 1.8 to 2.8. For $n_g = 1.9$ reflection into the zeroth diffraction order reaches the value of 1 with an accuracy of $10^{-8}$. High reflection into the zeroth diffraction order is possible due to the reduction of light reflection into higher diffraction orders, as described in Section~\ref{sec:impact-of-nc} of the main text and Section~\ref{sec:S_maths} in the Supplementary Materials.
 Maximum \textit{A} (Fig. \ref{fig:refl_map_S}b) reveals a broader $n_g$ range of above 0.99 reflectance into the zeroth diffraction order in comparison to maximum \textit{B}, ranging from 1.75 to 3.5. This is the limit for an ICG when assuming $n_c = 3.5$. In a very similar range from 1.8 to 3.5, total power reflectance is above $1 - 10^{-4}$, with the exception of the range from 2.8 to 3.25 where it reduces to 0.996.
 When tracking both maxima, several discontinuities in their geometrical parameters can be noted (see Fig. \ref{fig:refl_map_S}c). Starting from high $n_g$ values, the first is located at $n_g = 3.0$ for the \textit{A} maximum. This is a consequence of the \textit{A} maximum approaching the value of $\lambda/L$, where the third diffraction order appears. Other discontinuities in the geometrical parameters for maximum \textit{B} are at both limits of the $n_g$ range, from 2.1 to 2.2, where the maximum locates at the subwavelength limit at $L = \lambda$, as illustrated in the reflection map in Fig. \ref{fig:refl_map_S}g. The last two discontinuities in the parameters for maximum \textit{A} are present at the both ends of the $n_g$ range from 1.73 to 1.82, where \textit{A} maximum merges with the \textit{B} maximum. This range is indicated by two black vertical lines in Figs. \ref{fig:refl_map_S}a-c. The reflection map in Fig. \ref{fig:refl_map_S}e illustrates merging of both maxima in this $n_g$ range.
 Figures \ref{fig:refl_map_S}h and i are composed of zeroth diffraction order reflectance spectra of maximum \textit{B} and \textit{A} as a function of the grating refractive index ($n_g$). Figure \ref{fig:refl_map_S}i indicates the tendency of the width of reflection stopband (WRS, which we define as above 60\% of the reflection stopband) to narrow as $n_g$ reduces. The WRS of the \textit{B} maximum is significantly narrower with ``sharp'' feature in the proximity of the maximum. Therefore, the \textit{B} maximum can be considered as a possible narrowband filter while the \textit{A} maximum shows features that may be useful for mirrors in Fabry-Perot resonators. The discontinuities in the map are a consequence of the discontinuities in the grating parameters discussed in the context of Fig. \ref{fig:refl_map_S}c.

\newpage

\section{\label{sec:S_calc_real}Calculations for a real-world IP-DIP ICG}

\begin{figure*}[tbh]
\center\includegraphics[width=\textwidth]{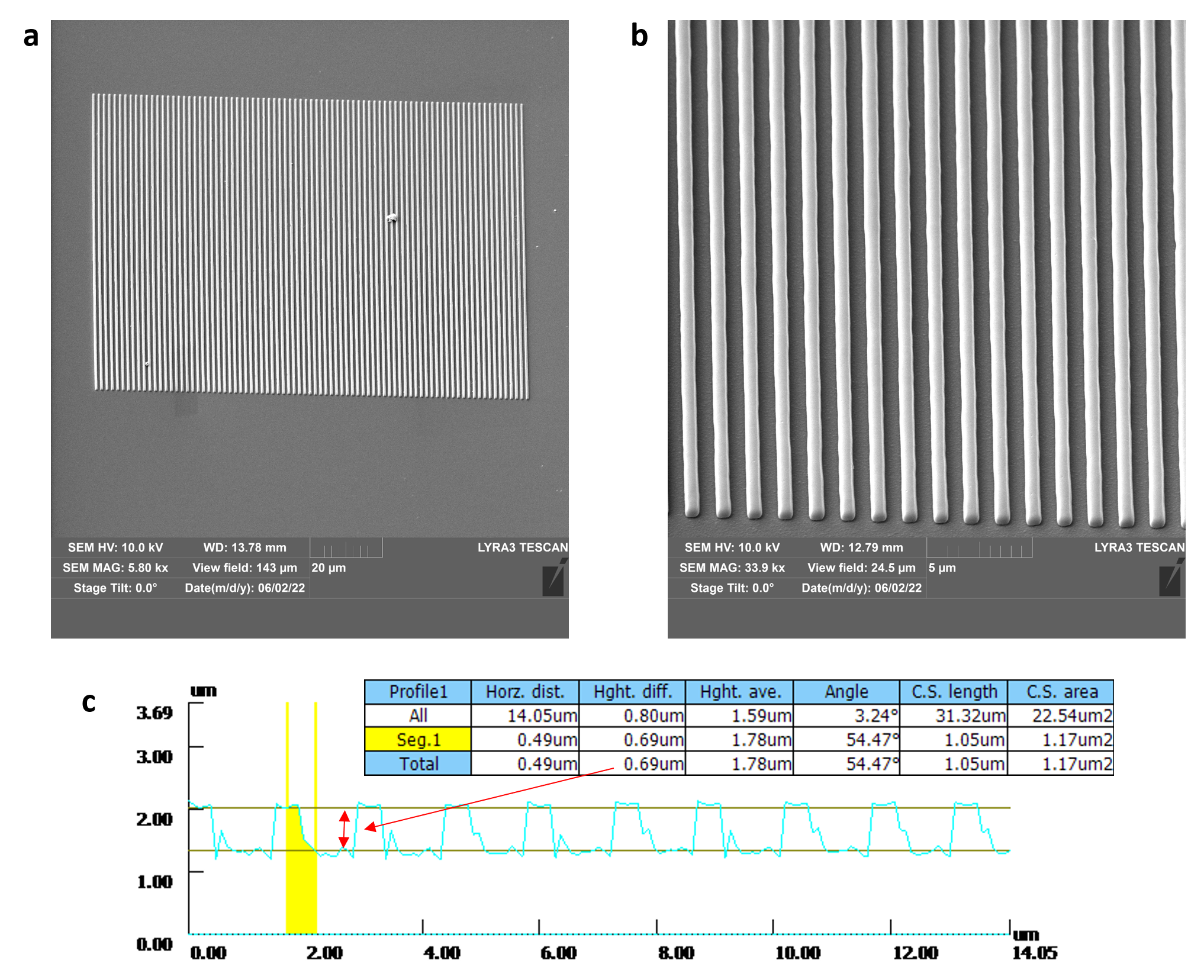}
\caption{\label{fig:SEM_balwan_real}SEM images of IP-Dip inverted refractive index contrast gratings (ICG): a), b) top-down view images of ICG stripes with different magnifications, images taken at an angle of 45 degrees to the plane of the wafer; c) profile of the ICG measured using a confocal microscope.}
\end{figure*}

Figures \ref{fig:SEM_balwan_real}a and \ref{fig:SEM_balwan_real}b show SEM images of the fabricated ICG. The cross-sectional shapes of the stripes exhibit a non-rectangular cross section, which we attribute to interference of the laser light in the vicinity of the substrate surface and to the high susceptibility of the two-photon absorption process to the laser power. In both images, the original scale is presented to enable determination of the lateral dimensions of the ICG.
In Fig. \ref{fig:SEM_balwan_real}c, results of profilometer measurements are shownm, indicating the very high repeatability of the stripe height.\todo{W opisie Fig. \ref{fig:SEM_balwan_real} wycięłam chwilowo "investigated by ???"}

\begin{figure*}[tbh]
\center\includegraphics[width=0.8\textwidth]{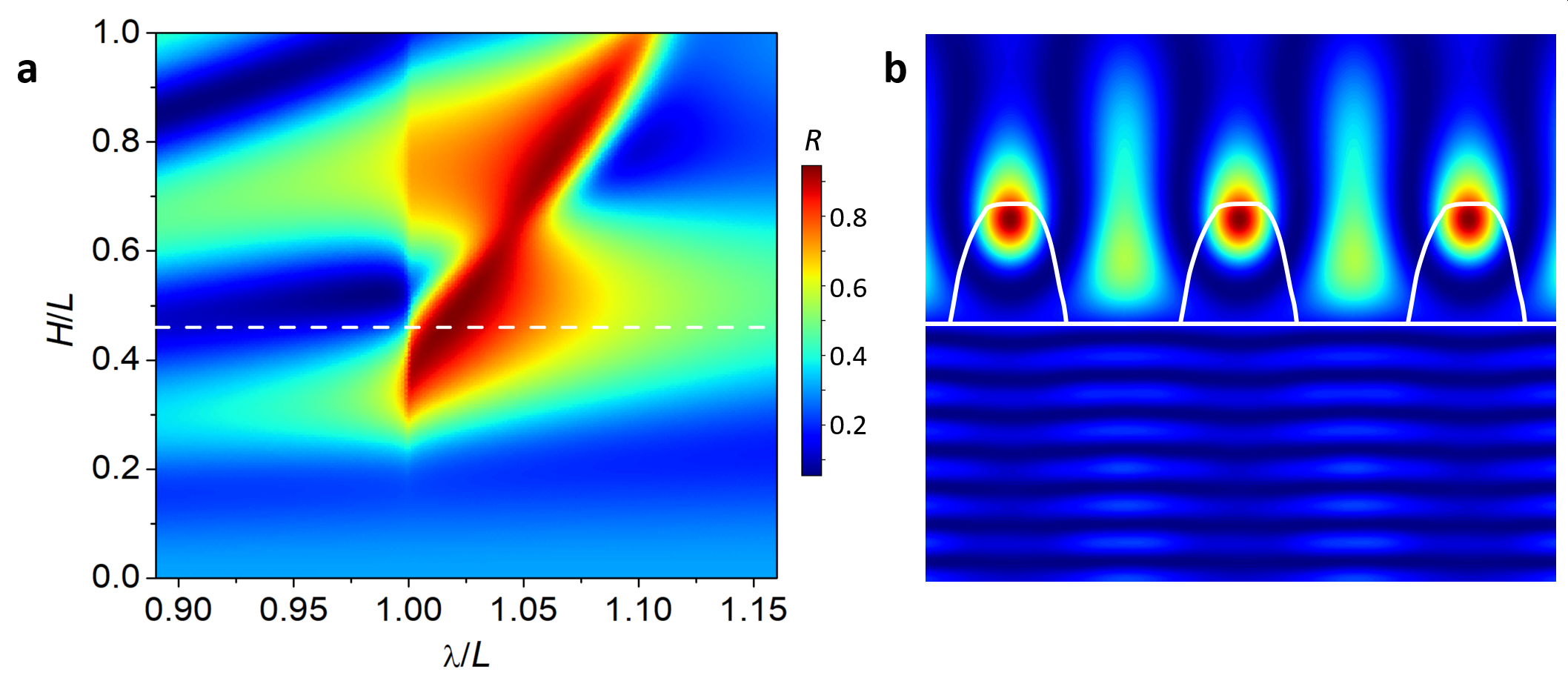}
\caption{\label{fig:balwan_cyfr} a) Calculated reflectance map for IP-Dip inverted refractive index contrast gratings (ICGs) for the cross section of the stripes illustrated in b), for $F = 0.45$ in the domain of the wavelength $\lambda$, and for the grating height $H$. The horizontal white dashed line represents $H/L = 0.46$. In b) is shown the distribution of optical field intensity within the IP-Dip ICG of the real-world cross section when illuminated by a plane wave at normal incidence from the cladding side.
}
\end{figure*}

Figure \ref{fig:balwan_cyfr}a shows a reflectance map of the ICG calculated for the experimental cross-sectional shape shown in  Fig. \ref{fig:SEM}c of the main text and illuminated by TE polarisation at normal incidence from the Si substrate side. The nonrectangular cross section of the ICG stripes is responsible for the more than 100 nm shift in the optimal $H$ toward smaller values compared to the rectangular cross section. Figure \ref{fig:balwan_cyfr}b illustrates the intensity of light incident from the substrate side, indicating a significant build-up of light intensity in the grating. The pattern resembles the distribution of light in the rectangular cross-section illustrated in Fig. \ref{fig:reflect_map_1}c.

\bibliographystyleS{naturemag}
\bibliographyS{supp}

\end{document}